\input harvmac
\input epsf
\input xy
\xyoption{all}
\noblackbox

\newcount\figno

\figno=0
\def\fig#1#2#3{
\par\begingroup\parindent=0pt\leftskip=1cm\rightskip=1cm\parindent=0pt
\baselineskip=11pt \global\advance\figno by 1 \midinsert
\epsfxsize=#3 \centerline{\epsfbox{#2}} \vskip 12pt
\centerline{{\bf Figure \the\figno :}{\it ~~ #1}}\par
\endinsert\endgroup\par}
\def\figlabel#1{\xdef#1{\the\figno}}
\def\pano{\par\noindent}
\def\smno{\smallskip\noindent}
\def\meno{\medskip\noindent}
\def\bigno{\bigskip\noindent}

\font\cmss=cmss10
\font\cmsss=cmss10 at 7pt

\def\rlx{\relax\leavevmode}
\def\inbar{\vrule height1.5ex width.4pt depth0pt}
\def\IC{\relax\,\hbox{$\inbar\kern-.3em{\rm C}$}}
\def\IR{\relax{\rm I\kern-.18em R}}
\def\IN{\relax{\rm I\kern-.18em N}}
\def\IP{\relax{\rm I\kern-.18em P}}
\def\ZZ{\rlx\leavevmode\ifmmode\mathchoice{\hbox{\cmss Z\kern-.4em Z}}
 {\hbox{\cmss Z\kern-.4em Z}}{\lower.9pt\hbox{\cmsss Z\kern-.36em Z}}
 {\lower1.2pt\hbox{\cmsss Z\kern-.36em Z}}\else{\cmss Z\kern-.4em Z}\fi}

\def\narrowplus{\kern -.04truein + \kern -.03truein}
\def\narrowminus{- \kern -.04truein}
\def\narrowminussub{\kern -.02truein - \kern -.01truein}

\def\o#1{\overline{#1}}

\def\ZX{{Z \hskip -8pt Z}}



\lref\rasreview{
C.~Angelantonj and A.~Sagnotti,
{\it Open Strings},
hep-th/0204089.
}

\lref\DoranVE{
C.~F.~Doran, M.~Faux and B.~A.~Ovrut,
{\it Four-dimensional N = 1 Super Yang-Mills Theory from an M
theory Orbifold},
hep-th/0108078.
}

\lref\FauxSP{
M.~Faux, D.~L\"ust and B.~A.~Ovrut,
{\it An M-theory Perspective on Heterotic K3 Orbifold Compactifications},
hep-th/0010087.
}

\lref\FauxDV{
M.~Faux, D.~L\"ust and B.~A.~Ovrut,
{\it Local Anomaly Cancellation, M-theory Orbifold and Phase-transitions},
Nucl.\ Phys.\ B {\bf 589} (2000) 269,
hep-th/0005251.
}

\lref\FauxHM{
M.~Faux, D.~L\"ust and B.~A.~Ovrut,
{\it Intersecting Orbifold Planes and Local
Anomaly Cancellation in  M-theory},
Nucl.\ Phys.\ B {\bf 554} (1999) 437,
hep-th/9903028.
}

\lref\rfaux{C.F.~Doran, M.~Faux, {\it
Intersecting Branes in M-Theory and Chiral Matter in Four Dimensions},
JHEP {\bf 0208} (2002) 024, hep-th/0207162.
}

\lref\rpradisi{G.~Pradisi, {\it Magnetized (Shift-)Orientifolds}, hep-th/0210088.
}

\lref\rkokora{C.~Kokorelis, {\it Deformed Intersecting D6-Brane GUTS I},
hep-th/0209202.
}

\lref\rkokorb{C.~Kokorelis, {\it Deformed Intersecting D6-Brane GUTS II},
hep-th/0210200.
}

\lref\rangles{M.~Berkooz, M.~R.~Douglas and R.~G.~Leigh, {\it Branes Intersecting
at Angles}, Nucl. Phys. B {\bf 480} (1996) 265, hep-th/9606139.
}

\lref\KleinVU{
M.~Klein, {\it Couplings in Pseudo-Supersymmetry},
Phys.Rev. {\bf D66} (2002) 055009,hep-th/0205300.
}

\lref\rKleinb{C.P. Burgess, E. Filotas, M. Klein, F. Quevedo,
{\it  Low-Energy Brane-World Effective Actions and Partial Supersymmetry Breaking},
hep-th/0209190.
}

\lref\rbailinb{D.~Bailin, G.V.~Kraniotis and A.~Love, {\it
Standard-like models from intersecting D5-branes},  hep-th/0210227.
}

\lref\rbailina{D.~Bailin, G.V.~Kraniotis and A.~Love, {\it
New Standard-like Models from Intersecting D4-Branes},  hep-th/0208103.
}

\lref\rrab{R.~Rabadan, {\it Branes at Angles, Torons, Stability and
Supersymmetry}, Nucl.\ Phys.\ B {\bf 620} (2002) 152, hep-th/0107036.
}

\lref\rbgkb{R.~Blumenhagen, L.~G\"orlich and B.~K\"ors,
{\it Supersymmetric Orientifolds in 6D with D-Branes at Angles},
Nucl.Phys. {\bf B569} (2000) 209, hep-th/9908130.
}

\lref\rbgkc{R.~Blumenhagen, L.~G\"orlich and B.~K\"ors, {\it
Supersymmetric 4D Orientifolds of Type IIA with D6-branes at Angles},
JHEP {\bf 0001} (2000) 040, hep-th/9912204.
}

\lref\rfhs{S.~F\"orste, G.~Honecker and R.~Schreyer, {\it Supersymmetric
$\ZZ_N \times \ZZ_M$ Orientifolds in 4-D with D-branes at Angles},
Nucl. Phys. B {\bf 593} (2001) 127, hep-th/0008250.
}

\lref\rfhstwo{S.~F\"orste, G.~Honecker and R.~Schreyer, {\it Orientifolds
with Branes at Angles}, JHEP {\bf 0106} (2001) 004, hep-th/0105208.
}

\lref\rbgklnon{R.~Blumenhagen, L.~G\"orlich, B.~K\"ors and D.~L\"ust,
{\it Noncommutative Compactifications of Type I Strings on Tori with Magnetic
Background Flux}, JHEP {\bf 0010} (2000) 006, hep-th/0007024.
}

\lref\rangela{R.~Blumenhagen and C.~Angelantonj,
{\it Discrete Deformations in Type I Vacua},
Phys.Lett. {\bf B473} (2000) 86, hep-th/9911190.
}

\lref\rbgotwo{work in progress}

\lref\IbanezDJ{
L.~E.~Ibanez,
{\it Standard Model Engineering with Intersecting Branes},
hep-ph/0109082.
}

\lref\rbgklmag{R.~Blumenhagen, L.~G\"orlich, B.~K\"ors and D.~L\"ust,
{\it Magnetic Flux in Toroidal Type I Compactification}, Fortsch. Phys. 49
(2001) 591, hep-th/0010198.
}

\lref\rba{C.~Angelantonj, R.~Blumenhagen, {\it Discrete Deformations in
Type I Vacua}, Phys. Lett. B {\bf 473} (2000) 86,
hep-th/9911190.
}

\lref\ras{C.~Angelantonj, A.~Sagnotti, {\it Type I
Vacua and Brane Transmutation}, hep-th/0010279.
}

\lref\raads{C.~Angelantonj, I.~Antoniadis, E.~Dudas, A.~Sagnotti, {\it Type I
Strings on Magnetized Orbifolds and Brane Transmutation},
Phys. Lett. B {\bf 489} (2000) 223, hep-th/0007090.
}

\lref\rbkl{R.~Blumenhagen, B.~K\"ors and D.~L\"ust,
{\it Type I Strings with $F$ and $B$-Flux}, JHEP {\bf 0102} (2001) 030,
hep-th/0012156.
}

\lref\rbgkl{R.~Blumenhagen, L.~G\"orlich, B.~K\"ors and D.~L\"ust,
{\it Asymmetric Orbifolds, Noncommutative Geometry and Type I
Vacua}, Nucl.\ Phys.\ B {\bf 582} (2000) 44, hep-th/0003024.
}

\lref\rbbkl{R.~Blumenhagen, V.~Braun, B.~K\"ors and D.~L\"ust,
{\it Orientifolds of K3 and Calabi-Yau Manifolds with Intersecting D-branes},
JHEP {\bf 0207} (2002) 026, hep-th/0206038.
}

\lref\rbbklb{R.~Blumenhagen, V.~Braun, B.~K\"ors and D.~L\"ust,
{\it The Standard Model on the Quintic}, hep-th/0210083.
}

\lref\rura{A.M.~Uranga,
{\it Local models for intersecting brane worlds}, hep-th/0208014.
}

\lref\rcvetica{M.~Cvetic, G.~Shiu and  A.~M.~Uranga,  {\it Three-Family
Supersymmetric Standard-like Models from Intersecting Brane Worlds}
Phys. Rev. Lett. {\bf 87} (2001) 201801,  hep-th/0107143.
}

\lref\rcveticb{M.~Cvetic, G.~Shiu and  A.~M.~Uranga,  {\it
Chiral Four-Dimensional N=1 Supersymmetric Type IIA Orientifolds from
Intersecting D6-Branes}, Nucl. Phys. B {\bf 615} (2001) 3, hep-th/0107166.
}

\lref\rott{R.~Blumenhagen, B.~K\"ors, D.~L\"ust and T.~Ott, {\it
The Standard Model from Stable Intersecting Brane World Orbifolds},
Nucl. Phys. B {\bf 616} (2001) 3, hep-th/0107138.
}

\lref\rottb{R.~Blumenhagen, B.~K\"ors, D.~L\"ust and T.~Ott, {\it
Intersecting Brane Worlds on Tori and Orbifolds}, hep-th/0112015.
}

\lref\rbonna{S.~F\"orste, G.~Honecker and R.~Schreyer, {\it
Orientifolds with Branes at Angles}, JHEP {\bf 0106} (2001) 004,
hep-th/0105208.
}

\lref\rbonnb{G.~Honecker, {\it Intersecting Brane World Models from
D8-branes on $(T^2 \times T^4/\ZZ_3)/\Omega R_1$ Type IIA Orientifolds},
JHEP {\bf 0201} (2002) 025, hep-th/0201037.
}

\lref\rqsusy{D.~Cremades, L.~E.~Ibanez and F.~Marchesano, {\it
SUSY Quivers, Intersecting Branes and the Modest Hierarchy Problem},
JHEP {\bf 0207} (2002) 009, hep-th/0201205.
}

\lref\rqsusyb{D.~Cremades, L.~E.~Ibanez and F.~Marchesano, {\it
     Intersecting Brane Models of Particle Physics and the Higgs Mechanism},
   JHEP {\bf 0207} (2002) 022,   hep-th/0203160.
}

\lref\AldazabalPY{
G.~Aldazabal, L.~E.~Ibanez and A.~M.~Uranga,
{\it Gauging Away the Strong CP Problem},
hep-ph/0205250.
}

\lref\rbachas{C.~Bachas, {\it A Way to Break Supersymmetry}, hep-th/9503030.
}

\lref\rafiruph{G.~Aldazabal, S.~Franco, L.~E.~Ibanez, R.~Rabadan, A.~M.~Uranga,
{\it Intersecting Brane Worlds}, JHEP {\bf 0102} (2001) 047, hep-ph/0011132.
}

\lref\rafiru{G.~Aldazabal, S.~Franco, L.~E.~Ibanez, R.~Rabadan, A.~M.~Uranga,
{\it $D=4$ Chiral String Compactifications from Intersecting Branes},
J.\ Math.\ Phys.\  {\bf 42} (2001) 3103, hep-th/0011073.
}

\lref\rimr{L.~E.~Ibanez, F.~Marchesano, R.~Rabadan, {\it Getting just the
Standard Model at Intersecting Branes},
JHEP {\bf 0111} (2001) 002, hep-th/0105155.
}

\lref\belrab{J.~Garcia-Bellido and R.~Rabadan, {\it Complex Structure Moduli
Stability in Toroidal Compactifications},
JHEP {\bf 0205} (2002) 042, hep-th/0203247.
}

\lref\rcim{D.~Cremades, L.~E.~Ibanez and F.~Marchesano, {\it
    Standard Model at Intersecting D5-branes: Lowering the String Scale},
   Nucl.Phys. {\bf B643} (2002) 93,  hep-th/0205074.
}

\lref\rkokoa{C.~Kokorelis, {\it GUT Model Hierarchies from Intersecting Branes},
JHEP {\bf 0208} (2002) 018, hep-th/0203187.
}

\lref\rkokob{C.~Kokorelis, {\it New Standard Model Vacua from Intersecting Branes},
JHEP {\bf 0209} (2002) 029, hep-th/0205147.
}

\lref\HoneckerDJ{
G.~Honecker,
{\it Non-supersymmetric Orientifolds with D-branes at Angles}, hep-th/0112174.
}

\lref\rcls{M.~Cvetic, P.~Langacker, and G.~Shiu, {\it
 Phenomenology of A Three-Family Standard-like String Model},
Phys.Rev. {\bf D66} (2002) 066004, hep-ph/0205252.
}

\lref\rclsb{M.~Cvetic, P.~Langacker, and G.~Shiu, {\it
 A Three-Family Standard-like Orientifold Model: Yukawa Couplings and Hierarchy},
Nucl.Phys. {\bf B642} (2002) 139, hep-ph/0206115.
}

\lref\rgkp{S.~B.~Giddings, S.~Kachru and J.~Polchinski, {\it
Hierarchies from Fluxes in String Compactifications},
hep-th/0105097.
}

\lref\rquev{C.P.~Burgess, L.E.~Ibanez and F.~Quevedo, {\it
Strings at the Intermediate Scale, or is the Fermi Scale Dual to the Planck Scale?},
Phys.Lett. {\bf B447} (1999) 257, hep-th/9810535.
}

\lref\DasguptaSS{
K.~Dasgupta, G.~Rajesh and S.~Sethi,
{\it M theory, orientifolds and G-flux},
JHEP {\bf 9908}, 023 (1999), hep-th/9908088.
}

\lref\rkst{S.~Kachru, M.~Schulz and S.~Trivedi, {\it
             Moduli Stabilization from Fluxes in a Simple IIB Orientifold},
              hep-th/0201028.
}

\lref\radd{N.~Arkani-Hamed, S.~Dimopoulos, and G.~Dvali, {\it The Hierarchy
Problem and New Dimensions at a Millimeter}, Phys. Lett. B {\bf 429} (1998)
263, hep-ph/9803315.
}

\lref\raadd{I.~Antoniadis, N.~Arkani-Hamed, S.~Dimopoulos, and G.~Dvali, {\it
New Dimensions at a Millimeter to a Fermi and Superstrings at a TeV},
Phys. Lett. B {\bf 436} (1998) 257, hep-ph/9804398.
}

\lref\ruranga{A.~M.~Uranga, {\it D-brane, Fluxes and Chirality},
hep-th/0201221.
}

\lref\ruram{A.~M.~Uranga, {\it Localized Instabilities at Conifolds},
hep-th/0204079.
}

\lref\rdouglasb{M.R.~ Douglas, {\it Enhanced Gauge Symmetry in M(atrix)
Theory}, JHEP 9707 (1997) 004, hep-th/9612126.
}

\lref\rdouglasa{M.R.~ Douglas and G.~Moore, {\it D-branes, Quivers, and ALE
Instantons}, hep-th/9603167.
}


\lref\rfonti{
G.~Aldazabal, A.~Font, L.~E.~Ibanez and G.~Violero,
{\it D = 4, N = 1, Type IIB Orientifolds},
Nucl.\ Phys.\ B {\bf 536} (1998) 29, hep-th/9804026.
}

\lref\DixonJC{
L.~J.~Dixon, J.~A.~Harvey, C.~Vafa and E.~Witten,
{\it Strings On Orbifolds 2},
Nucl.\ Phys.\ B {\bf 274} (1986) 285.
}

\lref\DixonJW{
L.~J.~Dixon, J.~A.~Harvey, C.~Vafa and E.~Witten,
{\it Strings On Orbifolds},
Nucl.\ Phys.\ B {\bf 261} (1985) 678.
}

\lref\BailinIE{
D.~Bailin, G.~V.~Kraniotis and A.~Love,
{\it Standard-like Models from Intersecting D4-branes},
Phys.\ Lett.\ B {\bf 530} (2002) 202,
hep-th/0108131.
}

\lref\rfractio{D.E.~Diaconescu and  J.~Gomis, {\it
Fractional Branes and Boundary States in Orbifold Theories},
JHEP {\bf 0010} (2000) 001, hep-th/9906242.
}

\lref\rangi{C.~Angelantonj, I.~Antoniadis, G.~D'Appollonio, E.~Dudas 
     and A.~Sagnotti, {\it Type I vacua with brane supersymmetry breaking},
Nucl.Phys. {\bf B572} (2000) 36, hep-th/9911081.
}

\lref\rrabklein{M.~Klein and R.~Rabadan, {\it
D=4, N=1 orientifolds with vector structure}, Nucl.Phys. {\bf B596} (2001) 197,
hep-th/0007087.
}


\lref\KakushadzeEG{
Z.~Kakushadze,
{\it On Four-dimensional N = 1 Type I Compactifications},
Nucl.\ Phys.\ B {\bf 535} (1998) 311, hep-th/9806008.
}

\lref\sagn{M.~Bianchi and A.~Sagnotti, {\it On the Systematics of Open String
Theories}, Phys. Lett. B {\bf 247} (1990) 517.
}

\lref\rgimpol{
E.~G.~Gimon and J.~Polchinski, {\it Consistency Conditions
for Orientifolds and D-Manifolds}, Phys.\ Rev.\ {\bf D54} (1996) 1667,
hep-th/9601038.
}


\lref\rbdlr{I.~Brunner, M.~R.~Douglas, A.~Lawrence and C.~R\"omelsberger, {\it
  D-branes on the Quintic}, JHEP 0008 (2000) 015, hep-th/9906200.
}

\lref\rav{M.~Aganagic and C.~Vafa, {\it
         Mirror Symmetry, D-Branes and Counting Holomorphic Discs},
          hep-th/0012041.
}

\lref\rglift{S.~Kachru and J.~McGreevy, {\it
M-theory on Manifolds of $G_2$ Holonomy and Type IIA Orientifolds},
JHEP {\bf 0106} (2001) 027, hep-th/0103223.
}

\lref\rkklma{S.~Kachru, S.~Katz, A.~Lawrence and J.~McGreevy, {\it
      Open String Instantons and Superpotentials}, Phys.Rev. {\bf D62} (2000)
      026001, hep-th/9912151.
}

\lref\rkklmb{S.~Kachru, S.~Katz, A.~Lawrence and J.~McGreevy, {\it
             Mirror Symmetry for Open Strings}, hep-th/0006047.
}

\lref\rhklm{S.~Hellermann, S.~Kachru, A.~Lawrence and J.~McGreevy, {\it
               Linear Sigma Models for Open Strings}, hep-th/0109069 .
}

\lref\rkachmca{S.~Kachru and  J.~McGreevy, {\it
            Supersymmetric Three-cycles and (Super)symmetry Breaking},
               Phys.Rev. {\bf D61} (2000) 026001,  hep-th/9908135.
}

\lref\rwittena{E.~Witten, {\it
            BPS Bound States Of D0-D6 And D0-D8 Systems In A B-Field},
               JHEP {\bf 0204} (2002) 012,  hep-th/0012054.
}

\lref\rsen{A.~Sen, {\it
          Stable Non-BPS Bound States of BPS D-branes },
               JHEP {\bf 9808} (1998) 010,  hep-th/9805019.
}

\lref\rgaber{M.R~Gaberdiel, {\it
          Lectures on Non-BPS Dirichlet branes},
       Class.Quant.Grav. {\bf 17} (2000) 3483,  hep-th/0005029.
}

\lref\BlumenhagenUA{
R.~Blumenhagen, B.~K\"ors, D.~L\"ust and T.~Ott,
{\it Hybrid Inflation in Intersecting Brane Worlds},
Nucl.Phys. {\bf B641} (2002) 235, hep-th/0202124.
}

\lref\AganagicGS{
M.~Aganagic and C.~Vafa,
{\it Mirror Symmetry, D-branes and Counting Holomorphic Discs}
hep-th/0012041.
}

\lref\refPradisi{
G.~Pradisi,
{\it Type I Vacua from Diagonal $\ZZ_3$-Orbifolds},
Nucl.\ Phys.\ B {\bf 575} (2000) 134,
hep-th/9912218.
}

\Title{\vbox{
 \hbox{HU--EP-02/46}
 \hbox{hep-th/0211059}}}
{\vbox{\centerline{Supersymmetric Intersecting Branes on the }
\vskip 0.3cm \centerline{Type IIA $T^6/\ZX_4$ Orientifold}
}}
\centerline{Ralph Blumenhagen{}, Lars G\"orlich{} and Tassilo Ott{} }
\bigskip\medskip
\centerline{ {\it Humboldt-Universit\"at zu Berlin, Institut f\"ur
Physik,}}
\centerline{\it Invalidenstrasse 110, 10115 Berlin, Germany}
\centerline{\tt e-mail:
blumenha, goerlich, ott@physik.hu-berlin.de}
\bigskip
\bigskip

\centerline{\bf Abstract}
\noindent
We study supersymmetric
intersecting D6-branes wrapping 3-cycles in the Type IIA
$T^6/\ZZ_4$ orientifold background. As a new feature, the 3-cycles
in  this orbifold space  arise  both from the untwisted and the
$\ZZ_2$ twisted sectors. We present an integral basis for the
homology lattice, $H_3(M,\ZZ)$, in terms of fractional 3-cycles,
for which the intersection form involves the Cartan matrix of
$E_8$. We show that these fractional D6-branes can be used to
construct supersymmetric brane configurations realizing a three
generation Pati-Salam model. Via brane recombination
processes preserving supersymmetry, this GUT model can be
broken down to a standard-like model.


\Date{11/2002}
\newsec{Introduction}

Intersecting brane world models have been the subject of elaborate
string model building for several years
\refs{\rbgklnon\raads\rbgklmag\ras\rafiru\rafiruph\rbkl\rimr\rbonna\rrab
\rott\rcvetica\rcveticb\BailinIE\IbanezDJ\rottb\HoneckerDJ\rbonnb\rqsusy
\BlumenhagenUA\rqsusyb\rkokoa\belrab\rcim\rkokob\AldazabalPY\rcls\rclsb
\KleinVU\rbbkl\rura\rbailina\rKleinb\rkokora\rbbklb\rpradisi\rkokorb-\rbailinb}.
The main new ingredient in these models is that they contain
intersecting D-branes and open strings in a consistent manner
providing  simple mechanisms to generate chiral fermions and  to
break supersymmetry \refs{\rbachas,\rangles}. Most attempts for
constructing realistic models were dealing with non-supersymmetric
configurations of D-branes, mainly because non-trivial, chiral
intersecting brane world models are not easy to find. It is known
for instance that flat factorizing D-branes on the six-dimensional
torus as well as on the $T^6/\ZZ_3$ orbifold can never give rise
to globally supersymmetric models except for the trivial
non-chiral configuration where all D6-branes are located on top of
the orientifold plane \refs{\rott}. Supersymmetric models clearly
have some advantages over the non-supersymmetric ones. From the
stringy point of view such models are stable, as not only the
Ramond-Ramond (R-R) tadpoles cancel but also the
Neveu-Schwarz-Neveu-Schwarz (NS-NS) tadpoles. From the
phenomenological point of view, since the gauge hierarchy problem
is solved by supersymmetry, one can work in  the conventional
scenarios with a large string scale close to the Planck  scale or
in an intermediate regime \rquev.
For an overview on other Type I constructions see \rasreview.

The only semi-realistic supersymmetric models that have been found
so far are defined in the $T^6/\ZZ_2\times \ZZ_2$ orientifold
background and were studied in a series of papers
\refs{\rcvetica,\rcveticb,\rcls,\rclsb,\rpradisi}. Besides their
phenomenological impact, Type IIA supersymmetric intersecting
brane worlds with orientifold six-planes and D6-branes are also
interesting from the stringy point of view, as they are expected
to lift to M-theory on singular $G_2$ manifolds \rglift.

The aim of this paper is to pursue the study of intersecting brane
worlds on orientifolds with a particular emphasis on the
systematic construction  of semi-realistic globally supersymmetric
configurations. Note, that without the orientifold projection
supersymmetric intersecting brane configurations do not exist, as
the overall tension always would be positive. Interestingly, from
the technical point of view, the $\ZZ_4$ orbifold involves some
new insights, as not all 3-cycles are inherited from the torus. In
fact, a couple of 3-cycles arise in the $\ZZ_2$ twisted sector
implying that this model contains so-called fractional D6-branes,
which have been absent in the  $\ZZ_2\times \ZZ_2$ and $\ZZ_3$
orbifolds. To treat these exceptional cycles accordingly, we will
make extensive use of the formalism developed in \rbbkl.

It will turn out that supersymmetric models in general can be
constructed in a straightforward way. But as in other model
building approaches, finding semi-realistic three generation
models turns out to be quite difficult. Fortunately, we will
finally succeed in constructing a globally  supersymmetric three
generation Pati-Salam model with gauge group $SU(4)\times
SU(2)_L\times SU(2)_R$ and the Standard Model matter in addition
to some exotic matter in the symmetric and antisymmetric
representation of the two $SU(2)$ gauge groups. In this paper, we
will mainly focus on the new and interesting string model building
aspects and leave a detailed investigation of the phenomenological
implications of the discussed models for future work.

This paper is organized as follows. In section 2 we review some of
the material presented in  \rbbkl\ about the general structure of
intersecting  brane worlds on Calabi-Yau manifolds. We will review
those formulas which will be extensively used in the rest of the
paper. In section 3 we start to investigate the $M=T^6/\ZZ_4$
orbifold and in particular derive an integral basis for the
homology group $H_3(M,\ZZ)$, for which the intersection form
involves the Cartan-matrix of the Lie-algebra $E_8$. The main
ingredient in the construction of such an integral basis will be
the physical motivated introduction of fractional D-branes which
also wrap around exceptional (twisted) 3-cycles in $M$. In section
4 we construct the orientifold models of Type IIA on the orbifold
$M$ and discuss the orientifold planes, the action of the
orientifold projection on the homology and the additional
conditions arising for supersymmetric configurations. In section 5
we construct as a first example a globally supersymmetric four
generation Pati-Salam model. Finally, in section 6 we elaborate on
a supersymmetric model with initial gauge symmetry $U(4)\times
U(2)^3\times U(2)^3$ and argue that by brane recombination it
becomes a supersymmetric three generation Pati-Salam model. By
using conformal field theory methods, for this model we determine
the chiral and also the massless non-chiral spectrum, which turns
out to provide Higgs fields in just the right representations in
order to break the model down to the Standard Model. At the end of
the paper we describe both the GUT breaking and the electroweak
breaking via brane recombination processes. We also make a
prediction for the Weinberg angle at the string scale.

\newsec{Intersecting Brane Worlds on Calabi-Yau spaces}

Before we present our new model, we would like to briefly
summarize some of the results presented in \rbbkl\ about Type IIA
orientifolds on smooth Calabi-Yau spaces. If the manifold admits
an anti-holomorphic involution $\o\sigma$, the combination
$\Omega\o\sigma$ is indeed a symmetry of the Type IIA model.
Taking the quotient with respect to this symmetry introduces an
orientifold six-plane into the background, which wraps a special
Lagrangian 3-cycle of the Calabi-Yau. In order to cancel  the
induced RR-charge, one introduces stacks of $N_a$ D6-branes which
are wrapped on 3-cycles $\pi_a$. Since under the action of
$\o\sigma$ such a 3-cycle, $\pi_a$,  is in general mapped to a
different 3-cycle, $\pi_a'$, one has to wrap the same number of
D6-branes on the latter cycle, too. The equation of motion for the
RR 7-form implies the RR-tadpole cancellation condition,
\eqn\tadhom{ \sum_a  N_a\, (\pi_a + \pi'_a)-4\, \pi_{O6}=0. } If
it is possible to wrap a connected smooth D-brane on such an
homology class,  the  stack of D6-branes supports a $U(N_a)$ gauge
factor. Note, that it is not a trivial question if in a given
homology class such a connected smooth manifold does exist.
However, as we will see in section 6 for special cases, there are
physical arguments ensuring that such smooth D-branes exist.

The Born-Infeld action provides an expression for the open string
tree-level scalar potential which by differentiation leads to an
equation for the NS-NS tadpoles \eqn\susy{ {V}=T_6\,
{e^{-\phi_{4}} \over M_s^3\sqrt{{\rm Vol({M})}}}
               \left( \sum_a  N_a \left( {\rm Vol}({\rm D}6_a) +
              {\rm Vol}({\rm D}6'_a) \right) -4\, {\rm Vol}({\rm O}6)\right) }
with the four-dimensional dilaton given by
$e^{-\phi_{4}}=M_s^3\sqrt{{\rm Vol({M})}}e^{-\phi_{10}}$ and $T_6$
denoting the tension of the $D6$-branes. By ${\rm Vol}({\rm
D}6_a)$ we mean the three dimensional internal volume of the
$D6$-branes. Generically, this scalar potential is non-vanishing
reflecting the fact that intersecting branes do break
supersymmetry. If the cycles are special Lagrangian (sLag) but
calibrated with respect to $3$-forms $\Re(e^{i\theta}\Omega_3)$
with different constant phase factors, the expression gets
simplified to \eqn\dbi{ {V}=T_6\, e^{-\phi_{4}} \left( \sum_a{N_a
\left| \int_{\pi_a} \widehat\Omega_3 \right|} +
 \sum_a{N_a \left| \int_{\pi'_a} \widehat\Omega_3 \right|}-
4 \left| \int_{\pi_{{\rm O}6}} \widehat\Omega_3 \right |\right) .
} In this case, all D6-branes preserve some supersymmetry but not
all of them the same. Models of this type have been discussed in
\refs{\rqsusy,\rqsusyb}. In the case of a globally supersymmetric
model, all 3-cycles are calibrated with respect to the same 3-form
as the O6-plane implying that the disc level scalar potential
vanishes due to the RR-tadpole condition \tadhom.

In \rbbkl\  it was argued and confirmed by many examples that the
chiral massless spectrum charged under the $U(N_1)\times
\ldots\times U(N_k)$ gauge group of a configuration of $k$
intersecting stacks of D6-branes can be computed from the
topological intersection numbers as shown in Table 1. \vskip 0.8cm
\vbox{ \centerline{\vbox{ \hbox{\vbox{\offinterlineskip
\def\tablespace{height2pt&\omit&&
 \omit&\cr}
\def\tablerule{\tablespace\noalign{\hrule}\tablespace}

\hrule\halign{&\vrule#&\strut\hskip0.2cm\hfill #\hfill\hskip0.2cm\cr
& Representation  && Multiplicity &\cr
\tablerule
& $[{\bf A_a}]_{L}$  && ${1\over 2}\left(\pi'_a\circ \pi_a+\pi_{{\rm O}6} \circ \pi_a\right)$   &\cr
\tablerule
& $[{\bf S_a}]_{L}$
     && ${1\over 2}\left(\pi'_a\circ \pi_a-\pi_{{\rm O}6} \circ
\pi_a\right)$   &\cr \tablerule & $[{\bf (\o N_a,N_b)}]_{L}$  &&
$\pi_a\circ \pi_{b}$ &\cr \tablerule & $[{\bf (N_a, N_b)}]_{L}$ &&
$\pi'_a\circ \pi_{b}$   &\cr }\hrule}}}} \centerline{ \hbox{{\bf
Table 1:}{\it ~~ Chiral spectrum in $d=4$}}} } \vskip 0.5cm
\noindent
Since in six dimensions the intersection number between two
3-cycles is anti-symmetric, the self intersection numbers do vanish
implying the absence of chiral fermions in the adjoint
representation. Negative intersection numbers correspond to chiral
fermions in the conjugate representations. Note, that if we want
to apply these formulas to orientifolds on singular toroidal
quotient spaces, the intersection numbers have to be computed in
the orbifold space and not simply in the ambient toroidal space.
After these preliminaries, we will discuss the $\ZZ_4$ orientifold in
the following sections.

\newsec{3-cycles in the  $\ZZ_4$ orbifold}

We consider Type IIA string theory compactified on the orbifold
background $T^6/\ZZ_4$, where the action of the $\ZZ_4$ symmetry,
$\Theta$, on the internal three complex coordinates reads
\eqn\actio{   z_1\to e^{\pi i \over 2}\, z_1, \quad
              z_2\to e^{\pi i \over 2}\, z_2, \quad
              z_3\to e^{-\pi i}\, z_3 }
with $z_1=x_1+i x_2$, $z_2=x_3+i x_4$ and $z_3=x_5+i x_6$.
This action preserves ${\cal N}=2$ supersymmetry in four dimensions
so that the orbifold describes a singular limit of a Calabi-Yau
threefold. The Hodge numbers of this threefold are given by
$h_{21}=7$ and $h_{11}=31$, where
1 complex and 5 K\"ahler moduli arise in the untwisted
sector. The $\Theta$ and $\Theta^3$ twisted sectors contain 16
$\ZZ_4$ fixed points giving rise to 16 additional K\"ahler moduli.
In the $\Theta^2$ twisted sector, there are 16 $\ZZ_2$ fixed points from
which 4 are also $\ZZ_4$ fixed points. The latter ones
contain 4 K\"ahler moduli whereas the remaining twelve $\ZZ_2$ fixed
points are organized in pairs under the $\ZZ_4$ action giving
rise to 6 complex and 6 K\"ahler moduli.
The fact that
the $\ZZ_2$ twisted sector contributes $h^{tw}_{21}=6$ elements to
the number of complex structure deformations and therefore contains
what might be called twisted 3-cycles, is the salient new feature
of this $\ZZ_4$ orbifold model as compared to the intersecting brane
world models studied so far.

Given this supersymmetric closed string background, we take
the quotient by the orientifold projection $\Omega\o\sigma$,
where $\o\sigma$ is an anti-holomorphic involution
$z_i\to  e^{i \phi_i} \o z_i$ of the manifold.
Note, that this orientifold model is not T-dual to
the $\ZZ_4$ Type IIB orientifold model studied first in \rfonti.
In the latter model there did not exist any
supersymmetric brane configurations cancelling all
tadpoles induced by the orientifold planes.
In fact, as was pointed out in \rbgkl\ our model is 
T-dual to a Type IIB orientifold on
an asymmetric $\ZZ_4$ orbifold space. 
Slightly different $\ZZ_4$ Type IIB orientifold models
were studied in \refs{\rangi,\rrabklein}.

Our orientifold projection breaks supersymmetry in the bulk to 
${\cal N}=1$ and introduces
an orientifold $O6$-plane located at the fixed point locus of
the anti-holomorphic involution.
The question arises if one can introduce D6-branes, generically not aligned
to the orientifold plane, in order to cancel the tadpoles induced by the
presence of the $O6$ plane.
The simplest such model where the D6-branes lie on top of the orientifold plane
has been investigated in \rbgkc.

\subsec{Crystallographic actions}

Before dividing Type IIA string theory by the discrete symmetries
$\ZZ_4$ and $\Omega\o\sigma$, we have to ensure that the torus
$T^6$ does indeed allow crystallographic actions of these symmetries.
For simplicity, we assume that $T^6$ factorizes as
$T^6=T^2\times T^2\times T^2$.
On the first two $T^2$s the $\ZZ_4$ symmetry enforces a rectangular
torus with complex structure $U=1$.
On each torus two different anti-holomorphic
involutions
\eqn\invol{\eqalign{  & {\bf A}: z_i\to  \o z_i \cr
                      & {\bf B}: z_i\to  e^{i {\pi\over 2}} \o z_i  }}
do exist. These two cases are shown in figure 1, where we have indicated the fixed point
set of the orientifold projection $\Omega\o\sigma$ \footnote{$^1$}{The same
distinction between the involutions ${\bf A}$ and ${\bf B}$ occurred
for the first time in the papers \refs{\rbgkb,\rangela,\rbgkc,\rfhs} .}.

\fig{Anti-holomorphic involutions}{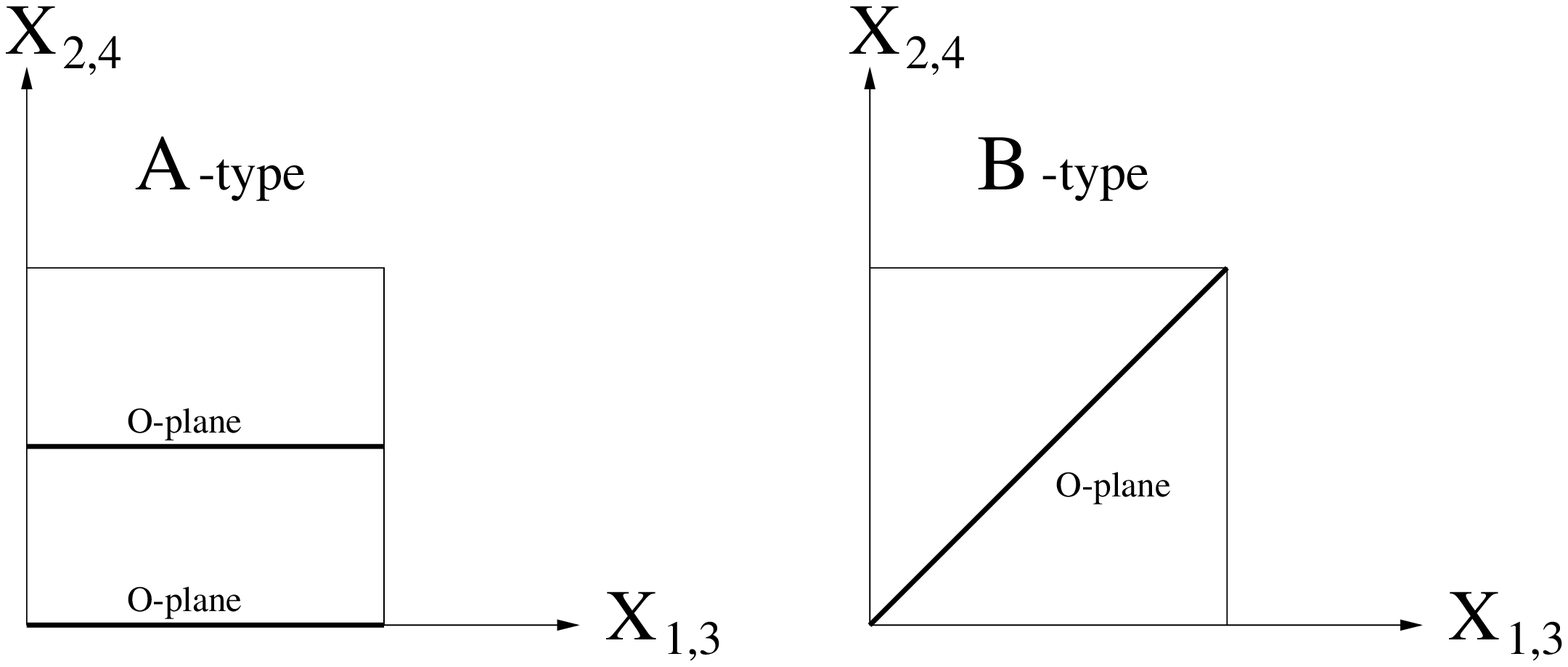}{10truecm}
\noindent
Since on the third torus the $\ZZ_4$ acts like a reflection, the
complex structure is unconstrained. But again there exist two different
kinds of involutions, which  equivalently correspond to  the two possible
choices of the orientation of the torus as shown in figure 2.
\fig{Orientations of the third $T^2$}{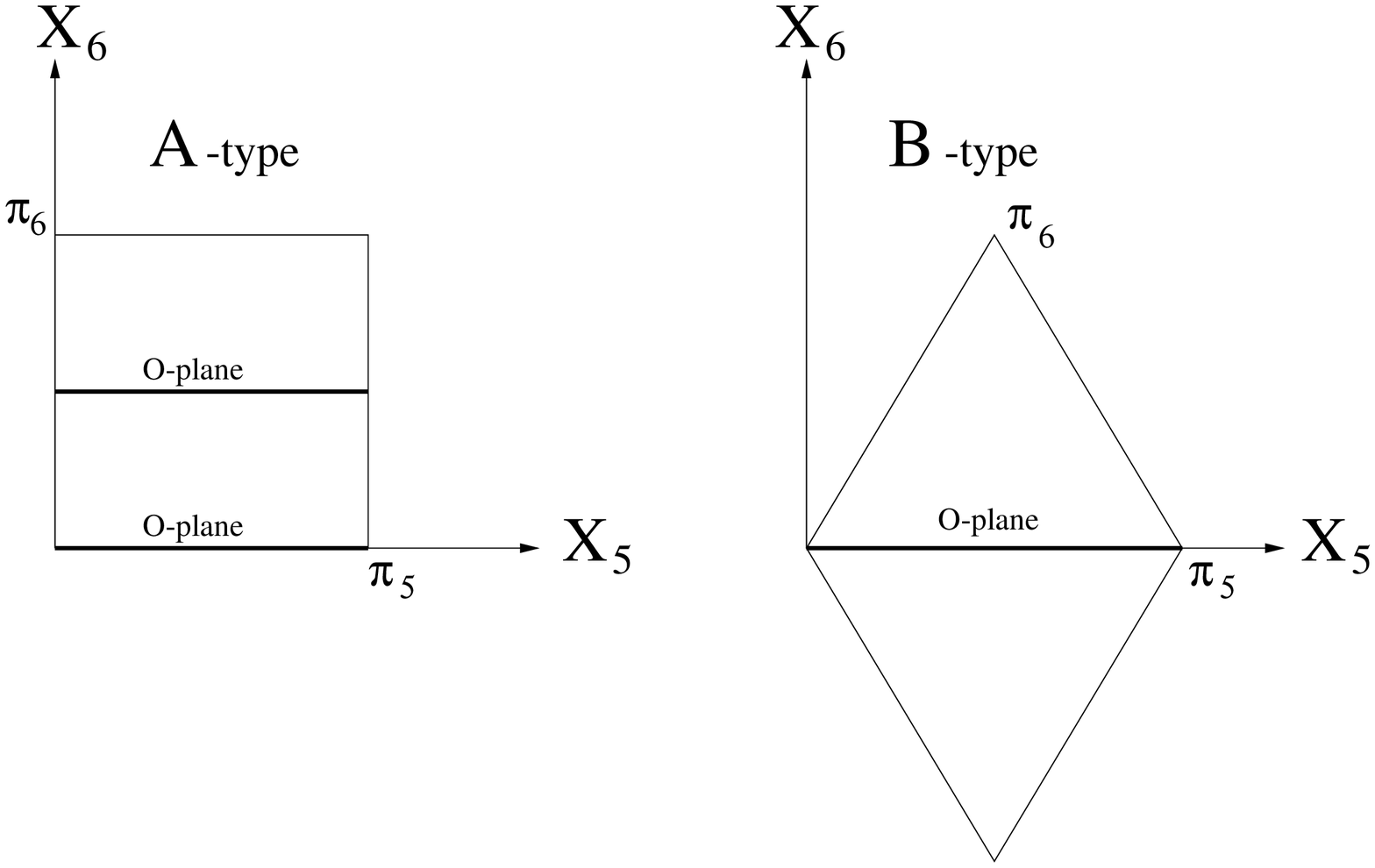}{12truecm}
\noindent
For the ${\bf A}$-torus the complex structure is given by $U=i U_2$
with $U_2$ unconstrained and for the ${\bf B}$-torus the complex structure
is given by $U={1\over 2}+i U_2$.
Therefore, by combining all possible choices of complex conjugations
we get eight possible orientifold models.
However, taking into account that the orientifold model on the $\ZZ_4$
orbifold does not only contain the orientifold planes related
to $\Omega\o\sigma$ but also the orientifold planes
related to $\Omega\o\sigma \Theta$, $\Omega\o\sigma \Theta^2$ and
$\Omega\o\sigma \Theta^3$, only four models
$\{${\bf AAA,ABA,AAB,ABB}$\}$ are actually different.

\subsec{A non-integral basis of 3-cycles}

In order to utilize  the formulas from section 2, we have to find
the independent 3-cycles on the $\ZZ_4$ orbifold space.
Since we already know that the third Betti number, $b_3=2+2 h_{21}$,
is equal to sixteen, we expect to find precisely this number
of independent 3-cycles.

One set of 3-cycles we get for free as they descend from the ambient space.
Consider the three-cycles  inherited from the torus $T^6$.
We call the two fundamental cycles on the torus $T^2_I$ ($I=1,2,3$)
$\pi_{2I-1}$ and $\pi_{2I}$ and moreover we define
the toroidal 3-cycles
\eqn\picyc{
\pi_{ijk} \equiv \pi_i\otimes\pi_j\otimes\pi_k.}
Taking orbits under the $\ZZ_4$ action, one can deduce the following
four $\ZZ_4$ invariant 3-cycles
\eqn\invcyc{\eqalign{
\rho_1 &\equiv 2 (\pi_{135}-\pi_{245} ),   \qquad
\bar{\rho}_1   \equiv 2 (\pi_{136}-\pi_{246} ) \cr
\rho_2 &\equiv 2 (\pi_{145}+\pi_{235} ), \qquad
\bar{\rho}_2  \equiv 2 (\pi_{146}+\pi_{236} )  .}}
The factor of two in \invcyc\ is due to the fact that
$\Theta^2$ acts trivially on the toroidal 3-cycles.
In order to compute the intersection form, we make use of the following
fact: if the 3-cycles $\pi^t_a$
on the torus are arranged in orbits of length $N$
under some  $\ZZ_N$ orbifold group, i.e.
\eqn\orbit{
\pi_a \equiv  \sum_{i=0}^{N-1} \Theta^i \pi^t_a ,
}
the intersection number between two such 3-cycles on the orbifold space
is given by
\eqn\inta{
\pi_a\circ\pi_b={1\over N} \left(\sum_{i=0}^{N-1} \Theta^i \pi^t_a
  \right) \circ \left(\sum_{j=0}^{N-1} \Theta^j \pi^t_b \right) .
}
Therefore, the intersection form for the four 3-cycles \invcyc\ reads
\eqn\form{   I_{\rho}=\bigoplus_{i=1}^2  \left(\matrix{ 0 & -2 \cr
                                                 2 & 0 \cr }\right) .}
The remaining twelve 3-cycles arise in the $\ZZ_2$ twisted
sector of the orbifold.
Since $\Theta^2$ acts non-trivially only onto the first two $T^2$,
in the $\ZZ_2$ twisted sector the sixteen $\ZZ_2$ fixed
points do appear as shown in figure 3.
\fig{Orbifold fixed points}{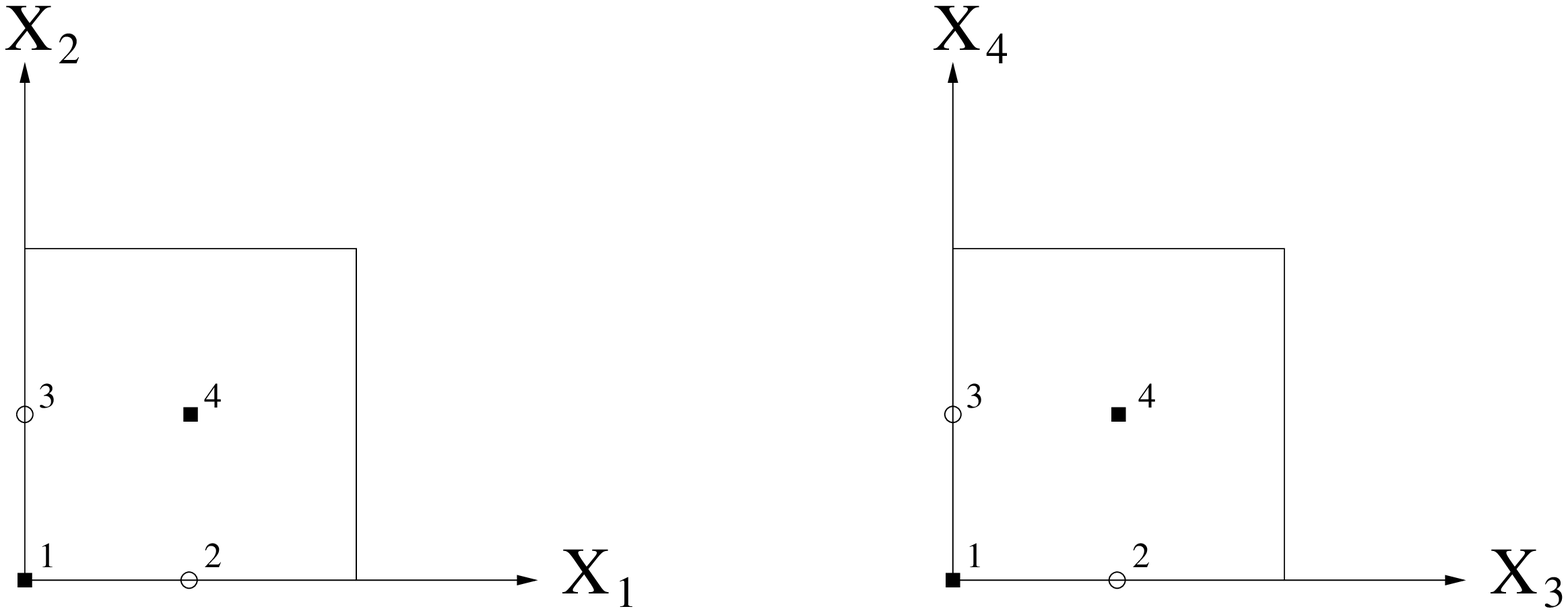}{12truecm}
\noindent
The boxes in the figure indicate the $\ZZ_2$ fixed points which are also
fixed under the $\ZZ_4$ symmetry.
After blowing up the
orbifold singularities, each of these fixed points gives rise an exceptional 2-cycle $e_{ij}$ with
the topology of $S^2$. These exceptional 2-cycles can be combined with
the two fundamental 1-cycles on the third torus to form what might
be called exceptional 3-cycles with the topology $S^2\times S^1$.
However, we have to take into account the $\ZZ_4$ action, which leaves
four fixed points invariant and arranges the remaining twelve in six
pairs.
Since the $\ZZ_4$ acts by reflection on the third torus,
its action on the exceptional cycles $e_{ij}\otimes \pi_{5,6}$
is
\eqn\actionthe{   \Theta\left(e_{ij}\otimes \pi_{5,6}\right)=-
                    e_{\theta(i)\theta(j)} \otimes \pi_{5,6} }
with
\eqn\actiontheb{   \theta(1)=1,\quad  \theta(2)=3,\quad \theta(3)=2,\quad
                      \theta(4)=4.}
Due to the minus sign in \actionthe\ the
invariant $\ZZ_4$ fixed points drop out and what remains are precisely
the twelve 3-cycles
\eqn\excyc{\eqalign{
\varepsilon_{1}       & \equiv (e_{12}-e_{13})\otimes\pi_5,  \qquad
\bar{\varepsilon}_{1}  \equiv (e_{12}-e_{13})\otimes\pi_6 \cr
\varepsilon_{2}       & \equiv (e_{42}-e_{43})\otimes\pi_5, \qquad
\bar{\varepsilon}_{2}  \equiv (e_{42}-e_{43})\otimes\pi_6 \cr
\varepsilon_{3}       & \equiv (e_{21}-e_{31})\otimes\pi_5, \qquad
\bar{\varepsilon}_{3}  \equiv (e_{21}-e_{31})\otimes\pi_6 \cr
\varepsilon_{4}       & \equiv (e_{24}-e_{34})\otimes\pi_5, \qquad
\bar{\varepsilon}_{4}  \equiv (e_{24}-e_{34})\otimes\pi_6 \cr
\varepsilon_{5}       & \equiv (e_{22}-e_{33})\otimes\pi_5, \qquad
\bar{\varepsilon}_{5}  \equiv (e_{22}-e_{33})\otimes\pi_6 \cr
\varepsilon_{6}       & \equiv (e_{23}-e_{32})\otimes\pi_5, \qquad
\bar{\varepsilon}_{6}  \equiv (e_{23}-e_{32})\otimes\pi_6. }}
Utilizing \inta\ the resulting intersection form is simply
\foot{As explained in hep-th/0502095, there was a wrong overall sign 
in an older version of this paper.}
\eqn\formex{   I_{\varepsilon}=\bigoplus_{i=1}^6  \left(\matrix{ 0 & 2 \cr
                                                 -2 & 0 \cr }\right) .}
These 3-cycles  lie in $H_3(M,\ZZ)$ but do not form an
integral basis of the free module since their intersection form is not
unimodular.

\subsec{An integral basis of 3-cycles}

The cycles which are missing so far are the ones corresponding
to what is called fractional D-branes \refs{\rdouglasa,\rdouglasb}.
In our context these are
D-branes wrapping only one-half times around the toroidal
cycles $\{\rho_1,\bar\rho_1,\rho_2,\bar\rho_2\}$ while
wrapping simultaneously  around some of the exceptional 3-cycles.
Therefore in the orbifold limit such branes are stuck at the
fixed points and one needs at least two such fractional
D-branes in order to form a brane which can be moved into the bulk.

To proceed, we need a rule of what combinations of toroidal and
exceptional cycles are allowed  for a fractional D-brane.
Such a rule can be easily gained from our physical intuition.
A D-brane wrapping for instance the toroidal cycle
${1\over 2}\rho_1$ can only wrap around those exceptional
3-cycles that correspond to the $\ZZ_2$ fixed points the flat D-brane
is passing through.  In our case, when the brane is lying along
the x-axis on the three $T^2$s, the allowed exceptional
cycles are $\{\varepsilon_1,\varepsilon_3,\varepsilon_5\}$.
Therefore, the total homological cycle the D-brane is wrapping on  can be
for instance
\eqn\frac{ \pi_a={1\over 2}\rho_1+{1\over 2}(
                \varepsilon_1+\varepsilon_3+\varepsilon_5) .}
The relative signs for the four different terms in \frac\
are still free parameters and
at the orbifold point do correspond to turning on a discrete Wilson line
along a longitudinal internal direction of the D-brane.
 Note, that this construction is completely analogous to
the construction of boundary states for fractional D-branes
\refs{\rsen,\rfractio,\rgaber}
carrying also a charge under some $\ZZ_2$ twisted sector states.

As an immediate consequences of this rule, only unbarred
respectively  barred cycles can be combined into fractional cycles,
as they wrap the same fundamental 1-cycle on the third $T^2$.
Apparently, the only non-vanishing intersection numbers are
between barred and unbarred cycles. Any unbarred fractional D-brane
can be expanded as \eqn\expa{  \pi_a=v_{a,1} \rho_1 + v_{a,2}
\rho_2 + \sum_{i=1}^6 v_{a,i+2}\,
         \varepsilon_i }
with half-integer valued coefficients $v_{a,i}$.
By exchanging the two fundamental cycles
on the third $T^2$, we can associate to it  a barred brane
\eqn\expb{ \o\pi_a=v_{a,1} \o\rho_1 + v_{a,2} \o\rho_2 - \sum_{i=1}^6 v_{a,i+2}\,
       \o\varepsilon_i }

with the same coefficients $v_{a,i+8}=v_{a,i}$ for $i\in\{1,\ldots,8\}$.
Using our rule we can construct all linear combinations
with ``self''-intersection number $\pi\circ \o\pi=-2$, where
we also have to keep in mind that the cycles form a lattice, i.e.
integer linear combinations of cycles are again cycles.
\bigno
In the following we list all the fractional
3-cycles with  ``self''-intersection number $\pi\circ \o\pi=-2$.
These cycles can be divided into 3 sets:

\item{a.)} $\{(v_1,v_2;v_3,v_4;v_5,v_6;v_7,v_8)\quad|\quad v_1+v_2=\pm 1/2,\,
                                   v_3+v_4=\pm 1/2,\,
                                  v_5+v_6=\pm 1/2,\,
                                  v_7+v_8=\pm 1/2;\,
 v_1+v_3+v_5+v_7 = 0\ {\rm mod}\ 1\}$. These combinations are
obtained by observing which fixed points
the flat branes parallel to the fundamental cycles do
intersect. These define $8\cdot 16=128$ different fractional 3-cycles.

\item{b.)} $\{(v_1,v_2 ;v_3,v_4;0,0;0,0),\,(v_1,v_2 ;0,0;v_5,v_6;0,0),\,
(v_1,v_2 ;0,0;0,0;v_7,v_8),$

\item{\phantom{b.)}} $ (0,0;v_3,v_4;v_5,v_6 ;0,0)
,\, (0,0;v_3,v_4;0,0;v_7,v_8),\, (0,0;0,0;v_5,v_6;v_7,v_8) \quad|
v_i \in {\pm 1/2} \} \}$. The first three kinds of cycles  are
again constructed from branes lying parallel to the x,y-axis on one
$T^2$ and stretching along the diagonal on the other $T^2$ .
The remaining  three kinds of cycles  arise from integer
linear combinations of the cycles introduced so far.
Thus, in total this yields $6\cdot 16 = 96 $ 3-cycles  in the second  set.

\item{c.)} $\{(v_1,v_2 ;v_3,v_4;v_5,v_6;v_7,v_8)\quad |
         \quad {\rm exactly\ one }\ v_i=\pm
  1\ {\rm , rest\ zero} \}$. Only the vectors with $v_1=\pm
  1$ or $v_2\pm1$ can be derived from untwisted branes. They are purely
  untwisted. The purely twisted ones again arise from linear
   combinations. This third set contains $2\cdot 8= 16$
  3-cycles.

\noindent Altogether there are 240 of such 3-cycles with
``self''-intersection number $-2$,  which intriguingly just
corresponds to the number of roots of the $E_8$ Lie algebra. Now,
it is easy to write a computer program searching for a basis among
these 240 cycles, so that the intersection form takes the
following form \eqn\eacht{   I=\left(\matrix{ 0 & C_{E_8} \cr
                              -C_{E_8} & 0 \cr}\right) }
where $C_{E_8}$ denotes the Cartan matrix of $E_8$
\eqn\cartan{   C_{E_8}=\left(\matrix{
-2  &   1 &   0 &  0 &  0 &  0 &  0 &   0 \cr
    1  &  -2 &   1 &  0 &  0 &  0 &  0 &   0 \cr
    0  &   1 &  -2 &  1 &  0 &  0 &  0 &   0 \cr
    0  &   0 &   1 & -2 &  1 &  0 &  0 &   0 \cr
    0  &   0 &   0 &  1 & -2 &  1 &  0 &   1 \cr
    0  &   0 &   0 &  0 &  1 & -2 &  1 &   0 \cr
    0  &   0 &   0 &  0 &  0 &  1 & -2 &   0 \cr
    0  &   0 &   0 &  0 &  1 &  0 &  0 &  -2 \cr }\right). }
One possible choice for the ``simple roots'' is
\eqn\simple{\eqalign{
 \vec{v}_1 &= {1\over 2 }(-1,\phantom{-}0, -1,\phantom{-}0,-1,\phantom{-}0,-1,
\phantom{-}0) \cr
 \vec{v}_2 &= {1\over 2 } (\phantom{-}1,\phantom{-}0,\phantom{-}1,\phantom{-}0,
\phantom{-}1,\phantom{-}0,-1,\phantom{-}0) \cr
 \vec{v}_3 &= {1\over 2 } (\phantom{-}1,\phantom{-}0, -1,\phantom{-}0,-1,
\phantom{-}0,\phantom{-}1,\phantom{-}0) \cr
 \vec{v}_4 &= {1\over 2 } (-1,\phantom{-}0,\phantom{-}1,\phantom{-}0,
\phantom{-}0,\phantom{-}1,\phantom{-}0,\phantom{-}1) \cr
\vec{v}_5 &= {1\over 2 } (\phantom{-}0,\phantom{-}1,-1,
\phantom{-}0,\phantom{-}1,\phantom{-}0,\phantom{-}0,-1) \cr
 \vec{v}_6 &= {1\over 2 } (\phantom{-}0,-1,\phantom{-}1,\phantom{-}0,-1,
  \phantom{-}0,\phantom{-}0,-1) \cr
\vec{v}_7 &= {1\over 2 } (\phantom{-}0,\phantom{-}1,\phantom{-}0,\phantom{-}1,
\phantom{-}0,-1,\phantom{-}0,\phantom{-}1) \cr
\vec{v}_8 &= {1\over 2 } (\phantom{-}0,-1, \phantom{-}0,-1,\phantom{-}0, -1,
 \phantom{-}0,\phantom{-}1). \cr }}
Since the Cartan matrix is unimodular, we indeed have constructed
an integral basis for the homology lattice $H_3(M,\ZZ)$. In the
following, it turns out to be more convenient to work with the
non-integral orbifold basis allowing also half-integer
coefficients. However, as we have explained not all such cycles
are part of $H_3(M,\ZZ)$, so we have to ensure each time we use
such fractional 3-cycles that they are indeed contained in the
unimodular lattice $H_3(M,\ZZ)$, i.e. that they are integer linear
combinations of the basis \simple.

\newsec{Orientifolds on the $\ZZ_4$ orbifold}

Equipped with the necessary information about the 3-cycles in the
$\ZZ_4$ toroidal orbifold, we can move forward and consider the
four inequivalent orientifold models in more detail.

\subsec{The O6-planes}

First, we have to determine the 3-cycle of the O6-planes.
Let us discuss this computation for the ${\bf ABB}$
model in some more detail, as this orientifold will
be of main interest for its potential to provide
semi-realistic standard-like models.

We have to determine the fixed point sets of the four
relevant orientifold projections $\{ \Omega\o\sigma, \Omega\o\sigma\Theta,
\Omega\o\sigma\Theta^2,\Omega\o\sigma\Theta^3\}$.
The results are listed in Table 2.
\vskip 0.8cm
\vbox{
\centerline{\vbox{
\hbox{\vbox{\offinterlineskip
\def\tablespace{height2pt&\omit&&
 \omit&\cr}
\def\tablerule{\tablespace\noalign{\hrule}\tablespace}

\hrule\halign{&\vrule#&\strut\hskip0.2cm\hfill #\hfill\hskip0.2cm\cr
& Projection  && fixed point set &\cr
\tablerule
& $\Omega\,\o\sigma$  && $2\, \pi_{135} + 2\, \pi_{145}  $   &\cr
\tablerule
& $\Omega\,\o\sigma\,\Theta$  && $2\, \pi_{145} + 2\, \pi_{245}- 4\, \pi_{146}- 4\, \pi_{246}$   &\cr
\tablerule
& $\Omega\, \o\sigma\, \Theta^2$  && $2\, \pi_{235} - 2\, \pi_{245} $   &\cr
\tablerule
& $\Omega\, \o\sigma\, \Theta^3$  && $-2\, \pi_{135} + 2\, \pi_{235}+ 4\,
                  \pi_{136}- 4\, \pi_{236}$   &\cr
}\hrule}}}}
\centerline{
\hbox{{\bf Table 2:}{\it ~~ O6-planes for {\bf ABB} model}}}
}
\vskip 0.5cm
\noindent
Adding up all contributions we get
\eqn\orid{\eqalign{  \pi_{O6}&=4\, \pi_{145} +4\, \pi_{235} +4\, \pi_{136}
                       -4\,\pi_{246} -4\, \pi_{146} -4\, \pi_{236} \cr
                      &=2\, \rho_2+2\, \o\rho_1 -2\, \o\rho_2 .\cr }}
Thus, only bulk cycles appear in $\pi_{O6}$ reflecting the fact
that in the conformal field theory the orientifold planes
carry only charge under untwisted R-R fields \refs{\rbgkb,\rbgkc}.
The next step is to determine the action of $\Omega\o\sigma$ on the
homological cycles. This can easily be done for the orbifold
basis. We find for the toroidal 3-cycles
\eqn\actbas{\eqalign{ &\rho_1\to \rho_2, \quad  \o\rho_1\to \rho_2- \o\rho_2 \cr
                      &\rho_2\to \rho_1, \quad  \o\rho_2\to \rho_1- \o\rho_1. \cr}}
For the exceptional cycles we require consistency with the exact 
conformal field theory computation in \rbgkc\  leading to to the following
action
%
\eqn\actbasex{\eqalign{ &\varepsilon_1\to +\varepsilon_1 \quad\quad
                        \o\varepsilon_1\to \varepsilon_1-\o\varepsilon_1 \cr
                        &\varepsilon_2\to +\varepsilon_2 \quad\quad
                        \o\varepsilon_2\to \varepsilon_2-\o\varepsilon_2 \cr
                        &\varepsilon_3\to -\varepsilon_3 \quad\quad
                        \o\varepsilon_3\to -\varepsilon_3+\o\varepsilon_3 \cr
                        &\varepsilon_4\to -\varepsilon_4 \quad\quad
                        \o\varepsilon_4\to  -\varepsilon_4+\o\varepsilon_4 \cr
                        &\varepsilon_5\to -\varepsilon_6  \quad\quad
                        \o\varepsilon_5\to -\varepsilon_6+\o\varepsilon_6 \cr
                        &\varepsilon_6\to -\varepsilon_5 \quad\quad
                        \o\varepsilon_6\to -\varepsilon_5+\o\varepsilon_5 .\cr}}

Consistently, the orientifold plane \orid\ is invariant under the $\Omega\o\sigma$
action. 
For the other three orientifold models,
the results for the $O6$ planes and the action of $\Omega\o\sigma$ on the
homology lattice can be found in Appendix A.
In principle, we have now provided all the information that is necessary to
build intersecting brane world models on the $\ZZ_4$ orientifold.
However, since we are particularly interested in supersymmetric models we need
to have control not only over topological data of the D6-branes but
over the nature of the sLag cycles as well.

\subsec{Supersymmetric cycles}

The metric at the orbifold point is flat up to some isolated
orbifold singularities. Therefore, flat D6-branes in a given
homology class are definitely special Lagrangian.
We restrict our D6-branes to be flat and factorizable
in the sense that they can be described by six wrapping numbers,
$(n_I,m_I)$ with $I=1,2,3$, along the fundamental toroidal cycles,
where for each $I$ the integers $(n_I,m_I)$ are relatively coprime.
Given such a bulk brane, one can compute the homology class
that it wraps expressed in the $\ZZ_4$ basis
\eqn\bulki{\eqalign{   \pi^{bulk}_a=&\left[ (n_{a,1}\,n_{a,2} -
                    m_{a,1}\,m_{a,2}) n_{a,3}\right]\, \rho_1 +
                    \left[(n_{a,1}\,m_{a,2} + m_{a,1}\,n_{a,2})
                   n_{a,3}\right]\, \rho_2 + \cr
                    &\left[(n_{a,1}\,n_{a,2} - m_{a,1}\,m_{a,2})
                    m_{a,3}\right]\, \o\rho_1 +
                    \left[(n_{a,1}\,m_{a,2} + m_{a,1}\,n_{a,2})
                m_{a,3}\right]\, \o\rho_2 . \cr}}
For the ${\bf ABB}$ orientifold,
 the condition that such a D6-brane preserves the same supersymmetry
as the orientifold plane is simply
\eqn\susygh{      \varphi_{a,1}+ \varphi_{a,2}+ \varphi_{a,3}={\pi\over 4}
                 \ {\rm mod}\ 2\pi}
with
\eqn\tangi{  \tan\varphi_{a,1}={m_{a,1}\over n_{a,1}}, \quad
              \tan\varphi_{a,2}={m_{a,2}\over n_{a,2}}, \quad
               \tan\varphi_{a,3}={U_2\, m_{a,3}\over n_{a,3}+{1\over 2} m_{a,3} } .}
Taking the $\tan(...)$  on both sides of equation \susygh\ we can reformulate
the supersymmetry condition in terms of wrapping numbers
(Note, that this only yields a necessary condition as  $\tan(...)$ is just periodic mod $\pi$.)
\eqn\susywar{  U_2={ \left(n_{a,3}+{1\over 2} m_{a,3}\right) \over m_{a,3}}
                  { \left(n_{a,1}\,n_{a,2} - m_{a,1}\,m_{a,2} - n_{a,1}\,m_{a,2} -
               m_{a,1}\,n_{a,2} \right)
           \over \left(n_{a,1}\,n_{a,2} - m_{a,1}\,m_{a,2} + n_{a,1}\,m_{a,2} +
            m_{a,1}\,n_{a,2} \right)}. }
Therefore, the complex structure of the third torus in general is
already fixed by one supersymmetric D-brane. In case one introduces
more D6-branes, one gets non-trivial conditions on the wrapping numbers
of these D-branes.
The supersymmetry conditions for the other three orientifold models
are summarized in Appendix B.

Working only with the bulk branes \bulki, the model building possibilities
are  very restricted. In particular, it seems to be impossible
to get large enough gauge groups to accommodate the Standard Model
gauge symmetry, $U(3)\times U(2)\times U(1)$,  of at least rank six.
One such supersymmetric model with only bulk branes and rank four
has been constructed in \rbbkl.
Now, to enlarge the number of possibilities, we also allow such
flat, factorizable branes to pass through $\ZZ_2$ fixed points and
split into fractional D-branes. Thus, according to our rule
we allow fractional D-branes wrapping the cycle
\eqn\fractio{   \pi_a^{frac}={1\over 2} \pi_a^{bulk} + {n_{a,3} \over 2}
                        \left[ \sum_{j=1}^6  w_{a,j} \varepsilon_j\right]
                  +  {m_{a,3 }\over 2}
                        \left[ \sum_{j=1}^6  w_{a,j} \o\varepsilon_j\right] }
with $w_{a,j}\in\{ 0,\pm 1\}$.  To make contact with the formerly introduced
coefficients $v_{a,j}$ , we define
\eqn\formercoef{ v_{a,j}={n_{a,3} \over 2} \, w_{a,j},\quad\quad
                 v_{a,j+8}={m_{a,3} \over 2} \, w_{a,j}}
for $j\in\{1,\ldots,8\}$.
In \fractio\ we have taken into account that the $\ZZ_2$ fixed points
all lie on the first two two-dimensional tori and that on the third torus
fractional D-branes do have  winding numbers along the two fundamental
1-cycles. Moreover, since $\varepsilon_j$
and $\o\varepsilon_j$ only differ by the cycle on the third torus,
their coefficients in \fractio\ must indeed be equal.

These fractional D6-branes do correspond to the following boundary states
in the conformal field theory of the $T^6\over \ZZ_4$ orbifold model
\eqn\bound{\eqalign{
       |D^f;(n_I,m_I),\alpha_{ij}\rangle=&
       {1\over 4\sqrt 2} \left(\prod_{j=1}^2 \sqrt{n_j^2+m_j^2}\right)\,
           \sqrt{n_3^2+n_3 m_3 +{\textstyle{m_3^2\over 2}}} \,
          \biggl(
     \bigl|D;(n_I,m_I)\bigr\rangle_U + \cr
    &\phantom{dfssssssssdsdfadssdddsssssadsadssdfsdfsdffsd}
             \bigl|D;\Theta(n_I,m_I)\bigr\rangle_U\biggr)+\cr
      &{1\over 2\sqrt 2} \,
           \sqrt{n_3^2+n_3 m_3 +\textstyle{{m_3^2\over 2}}}\, \biggl(
     \sum_{i,j=1}^{4} \alpha_{ij} \bigl|D;(n_I,m_I),e_{ij}\bigr\rangle_T + \cr
    &\phantom{df{1\over 2\sqrt 2} \,
           \sqrt{n_3^2+n_3 m_3 +\textstyle{{m_3^2\over 2}}} }
  \sum_{i,j=1}^{4} \alpha_{ij} \bigl|D;\Theta(n_I,m_I),\Theta(e_{ij})
                  \bigr\rangle_T \biggr). \cr }}
In the schematic form of the boundary state \bound\ there are contributions
from both the untwisted and the $\ZZ_2$ twisted sector and we have taken
the orbit under the $\ZZ_4$ symmetry $\Theta$ with the following action
on the winding numbers
\eqn\windact{ \Theta(n_{1,2},m_{1,2})=(-m_{1,2}, n_{1,2}),\quad\quad
               \Theta(n_{3},m_{3})=-(n_3, m_{3}) }
implying that $\Theta^2$ acts like the identity on the boundary
states. This explains why only two and not four untwisted boundary
states do appear in \bound. Note, that in the sum over the $\ZZ_2$
fixed points, for each D6-brane precisely  four coefficients take
values $\alpha_{ij}\in\{-1,+1\}$ and the remaining ones are
vanishing. The $\alpha_{ij}$ are of course directly related to the
coefficients $w_i$ appearing in the description of the
corresponding fractional 3-cycles. For the interpretation of these
coefficients $\alpha_{ij}$, one has to remember that changing the
sign of $\alpha_{ij}$ corresponds to turning on a discrete $\ZZ_2$
Wilson line along one internal direction of the brane
\refs{\rsen,\rgaber}. The action of $\Theta$ on the twisted sector
ground states $e_{ij}$ is the same as in \actionthe. The
elementary boundary states like $|D;(n_I,m_I)\rangle_U$ are the
usual ones for flat $D6$ brane with wrapping numbers $(n_I,m_I)$
on $T^6=T^2\times  T^2\times T^2$ and can be found in Appendix C.
The important normalization factors in \bound\ are fixed by the
Cardy condition, stating that the result for the annulus partition
function must coincide for  the loop and the tree channel
computation.

Since  the brane and its $\ZZ_4$ image only break the
supersymmetry down to ${\cal N}=2$ , one gets a ${\cal N}=2$
$U(N)$ vector multiplet on each stack of fractional D-branes. The
scalars in these vector multiplets correspond to the position of
the D6-brane on the third $T^2$ torus, which is still an open
string modulus.

Coming back to the homology cycles, following our general rule for
fractional branes imposes further constraints on the
coefficients because only those exceptional cycles
are allowed to contribute which are intersected by the flat
D-brane. The only allowed exceptional 3-cycles are summarized in
Table 3, depending on the wrapping numbers of the first two tori
$T^2$. \vskip 0.8cm \vbox{ \centerline{\vbox{
\hbox{\vbox{\offinterlineskip
\def\tablespace{height2pt&\omit&&\omit&&\omit&&
 \omit&\cr}
\def\tablerule{\tablespace\noalign{\hrule}\tablespace}

\hrule\halign{&\vrule#&\strut\hskip0.2cm\hfill #\hfill\hskip0.2cm\cr
&   && $n_1$ odd, $m_1$ odd && $n_1$ odd, $m_1$ even && $n_1$ even, $m_1$ odd  &\cr
\tablerule
& $n_2$ odd &&    && $\varepsilon_3$,  $\varepsilon_4$ &&   $\varepsilon_3$, $\varepsilon_4$ &\cr
& $m_2$ odd &&    && $\varepsilon_5$,  $\varepsilon_6$ &&   $\varepsilon_5$, $\varepsilon_6$ &\cr
\tablerule
& $n_2$ odd &&  $\varepsilon_1$,  $\varepsilon_2$  && $\varepsilon_1$,
       $\varepsilon_3$, $\varepsilon_5$  && $\varepsilon_1$,
       $\varepsilon_3$, $\varepsilon_6$   &\cr
& $m_2$ even && $\varepsilon_5$,  $\varepsilon_6$  && $\varepsilon_1$,
       $\varepsilon_4$, $\varepsilon_6$  && $\varepsilon_1$,
       $\varepsilon_4$, $\varepsilon_5$  &\cr
&  &&  && $\varepsilon_2$,
       $\varepsilon_3$, $\varepsilon_6$  && $\varepsilon_2$,
       $\varepsilon_3$, $\varepsilon_5$  &\cr
&  &&  && $\varepsilon_2$,
       $\varepsilon_4$, $\varepsilon_5$  && $\varepsilon_2$,
       $\varepsilon_4$, $\varepsilon_6$  &\cr
\tablerule
& $n_2$ even &&  $\varepsilon_1$,  $\varepsilon_2$  && $\varepsilon_1$,
       $\varepsilon_3$, $\varepsilon_6$  && $\varepsilon_1$,
       $\varepsilon_3$, $\varepsilon_5$   &\cr
& $m_2$ odd && $\varepsilon_5$,  $\varepsilon_6$  && $\varepsilon_1$,
       $\varepsilon_4$, $\varepsilon_5$  && $\varepsilon_1$,
       $\varepsilon_4$, $\varepsilon_6$  &\cr
&  &&  && $\varepsilon_2$,
       $\varepsilon_3$, $\varepsilon_5$  && $\varepsilon_2$,
       $\varepsilon_3$, $\varepsilon_6$  &\cr
&  &&  && $\varepsilon_2$,
       $\varepsilon_4$, $\varepsilon_6$  && $\varepsilon_2$,
       $\varepsilon_4$, $\varepsilon_5$  &\cr }\hrule}}}}
\centerline{ \hbox{{\bf Table 3:}{\it ~~ allowed exceptional
cycles}}} } \vskip 0.5cm \noindent
At first glance, there is a
mismatch between the number of parameters describing a 3-cycle and
the corresponding  boundary state. For each D6-brane there are
three non-vanishing parameters $w_i$ but four $\alpha_{ij}$.
However, a flat fractional brane and its $\ZZ_4$ image always
intersect in precisely one $\ZZ_4$ fixed point times a circle on
the third $T^2$.

Since $\Theta$ acts on this fixed locus  with a minus sign, this
twisted sector effectively drops out of the boundary state \bound.
A different way of saying this is that at the intersection between
the brane and its $\ZZ_4$ image, there lives a hypermultiplet,
$\Phi_{adj}$, in the adjoint representation. Since it is an
${\cal N}=2$ supermultiplet, there exists a flat direction in the
D-term potential corresponding to the recombination of the
two branes into a single brane. This single brane of course no longer
runs to the $\ZZ_4$ invariant fixed point.
This brane recombination process is depicted in figure 4.

\fig{Recombined branes}{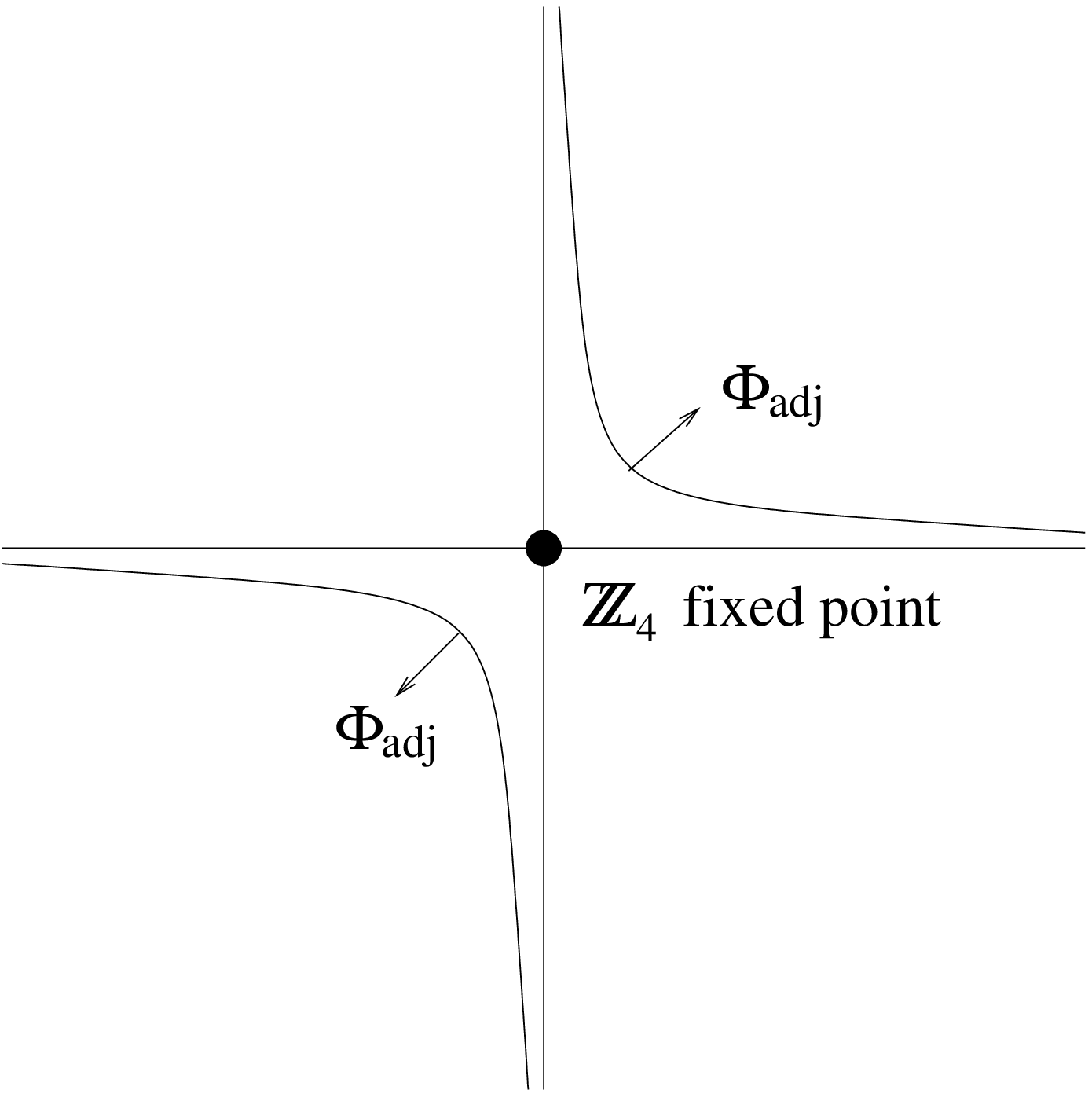}{8truecm}
\noindent

A non-trivial test for our
considerations is the condition that a fractional brane \fractio\
transformed to the $E_8$-basis must have  integer
coefficients. To see this, we write the $8\times8$ matrix \cartan\
and a second identical copy as the two diagonal blocks of a
$16\times16$ matrix, and then act with the inverse of the
transposed matrix onto a general vector \fractio . Then we have
to investigate the different cases according to Table 3
separately. For instance for the case $n_1$ odd, $n_2$ odd, $m_1$
even, $m_2$ odd and fractional cycles $\varepsilon_3$,
$\varepsilon_4$ with signs $w_{3}$, $w_{4}$ respectively, we
substitute $m_1=2k_1$
and  obtain the following vector in the $E_8$-basis: \eqn\proofinteger{\bigg[
\Big({1\over 2 }(n_{1} m_{2}-w_{3})+k_{1} n_{2}\Big)n_{3},\
 \Big({1 \over 2 }(n_{1} n_{2}-w_{3})-k_{1} m_{2}+n_{1} m_{2}+2
k_{1} n_{2}\Big)n_{3},\  ...\bigg]} Already for the first two
components we can see what generally happens for all cases and
components: since $n_1$, $n_2$, $m_2$ and $w_{3}$ are
non-vanishing and because products of odd numbers are also odd, just sums
and differences of two odd numbers occur and these are always even or
zero and therefore can be divided by 2 and still lead to
integer coefficients.
Having defined  a well understood set of supersymmetric
fractional D6-branes, we are now in
the position to search  for phenomenologically
interesting supersymmetric intersecting brane worlds.

\newsec{A four generation supersymmetric Pati-Salam model}

In this section we present the construction of a semi-realistic
supersymmetric intersecting brane world model. This provides  an
application of the formalism developed in the previous sections.
It turns out that the ${\bf ABB}$ orientifold model is the most
appropriate one for doing this. Using the fractional D6-branes
introduced in the last section, one finds that by requiring that
no (anti-)symmetric representations of the
$U(N_a)$ gauge groups do appear, only very few sufficiently small mutual
intersection numbers arise. For the ${\bf ABB}$ model with the
complex structure of the last torus being $U_2=1$, an extensive
computer search reveals that essentially only mutual intersection
numbers $(\pi_a\circ \pi_b,\pi'_a\circ \pi_b)=(0,0),(\pm 2,\mp
2)$ are possible. Even with these intersection numbers it is
possible to construct a four generation
supersymmetric Pati-Salam model with initial gauge group
$U(4)\times U(2)\times U(2)$. A typical model of this sort can be
realized by the following three stacks of D6-branes 
\vskip 0.8cm
\vbox{ \centerline{\vbox{ \hbox{\vbox{\offinterlineskip
\def\tablespace{height2pt&\omit&&\omit&&
 \omit&\cr}
\def\tablerule{\tablespace\noalign{\hrule}\tablespace}

\hrule\halign{&\vrule#&\strut\hskip0.2cm  #\hfill\hskip0.2cm\cr
& stack  && $(n_I,m_I)$  && homology cycle &\cr
\tablerule
& U(4) stack && $(-1,0;1,1;-1,0)$ && $\pi_1={1\over 2}\left(
          \rho_1+\rho_2+\varepsilon_5+\varepsilon_6\right)$ & \cr
& &&  && $\pi'_1={1\over 2}\left(
          \rho_1+\rho_2-\varepsilon_5-\varepsilon_6\right)$ & \cr
\tablerule
& U(2) stack && $(0,1;-1,-1;-1,2)$ && $\pi_2=
        {1\over 2}\left(
          -\rho_1+\rho_2+2\o\rho_1-2\o\rho_2+\varepsilon_5+\varepsilon_6-2
           \o\varepsilon_5-2\o\varepsilon_6 \right)$ & \cr
& &&  && $\pi'_2={1\over 2}\left(
          -\rho_1+\rho_2+2\o\rho_1-2\o\rho_2+\varepsilon_5+\varepsilon_6-2
           \o\varepsilon_5-2\o\varepsilon_6 \right)$ &\cr
\tablerule
& U(2) stack && $(-1,0;1,-1;1,-2)$ && $\pi_3=
        {1\over 2}\left(
          -\rho_1+\rho_2+2\o\rho_1-2\o\rho_2-\varepsilon_5-\varepsilon_6+2
           \o\varepsilon_5+2\o\varepsilon_6 \right)$ & \cr
& &&  && $\pi'_3={1\over 2}\left(
          -\rho_1+\rho_2+2\o\rho_1-2\o\rho_2-\varepsilon_5-\varepsilon_6+2
           \o\varepsilon_5+2\o\varepsilon_6 \right)$ &\cr
}\hrule}}}}
\centerline{
\hbox{{\bf Table 4:}{\it ~~ D6-branes for a 4 generation PS-model}}}
}
\noindent
Computing the intersection numbers for these D6-branes and using
the general formula for the chiral massless spectrum, one gets
the massless modes shown in Table 5.
\vskip 0.8cm
\vbox{
\centerline{\vbox{
\hbox{\vbox{\offinterlineskip
\def\tablespace{height2pt&\omit&&
 \omit&\cr}
\def\tablerule{\tablespace\noalign{\hrule}\tablespace}

\hrule\halign{&\vrule#&\strut\hskip0.2cm \hfill
#\hfill\hskip0.2cm\cr & n  && $SU(4)\times SU(2)\times SU(2)\times
U(1)^3$  &\cr \tablerule & 2 && $(4,2,1)_{(1,-1,0)}$ &\cr
\tablerule & 2 && $(4,2,1)_{(1,1,0)}$ &\cr \tablerule & 2 &&
$(\o{4},1,2)_{(-1,0,1)}$ &\cr \tablerule & 2 &&
$(\o{4},1,2)_{(-1,0,-1)}$ &\cr }\hrule}}}} \centerline{ \hbox{{\bf
Table 5:}{\it ~~ Chiral spectrum for 4 generation PS-model}}} }
\vskip 0.5cm \noindent Here we have normalized  as usual the gauge
fields in  the diagonal $U(1)_a\subset U(N_a)$ sub-algebras as
\eqn\normi{   A^\mu_{U(1)_a}={1\over N_a}\,{\rm Tr}
\left(A^\mu_{U(N_a)}\right) .} Note, that all non-abelian gauge
anomalies are canceled. Adding up all homological cycles, one
finds that the RR-tadpole cancellation condition \tadhom\ is
indeed satisfied. A nice check is whether the NS-NS tadpole
cancellation condition \susy\ is also satisfied, as it should be
for a globally supersymmetric configuration. For the contribution
of the O6-plane to the scalar potential, one finds \eqn\tens{
{\cal V}_{O6}=-T_6\, e^{-\phi_4} 16 \sqrt{2}\left(
                    {1\over \sqrt{U_2}} +2 \sqrt{U_2} \right) ,}
whereas the three stacks of D6-branes give
\eqn\tenso{\eqalign{
          {\cal V}_{1}&=T_6\, e^{-\phi_4} 16 \sqrt{2} {1\over \sqrt{U_2}} \cr
          {\cal V}_{2,3}&=T_6\, e^{-\phi_4} 16 \sqrt{2} \sqrt{U_2} .\cr}}

We see that the scalar potential vanishes for all values
of the complex structure $U_2$ of the third torus.
Thus, the disc level scalar potential indeed vanishes and
we have constructed a globally supersymmetric
intersecting brane world model with gauge group $U(4)\times
U(2)\times U(2)$.

\subsec{Green-Schwarz mechanism}

Computing in the usual way the mixed $U(1)_a-SU(N_b)^2$ anomalies,
one confirms the general result derived in \rbbkl\ \eqn\mixed{
A_{ab}={N_a\over 4}(-\pi_a+\pi'_a)\circ(\pi_b+\pi'_b) .} In our
example there is only one anomalous $U(1)$ while $U(1)_2$ and
$U(1)_3$ are anomaly-free. This anomaly is canceled by some
generalized Green-Schwarz mechanism involving the axionic
couplings from the Chern-Simons terms in the effective action on
the D6-branes \eqn\gsmech{  S^F_{CS}=  \sum_{i=1}^{b_3}    \int
d^4 x\,
              N_a\,  (v_{a,i} - v'_{a,i} )\, B_i\wedge F_a   }
and
\eqn\gsmechb{  S^{F\wedge F}_{CS}=  \sum_{i=1}^{b_3}    \int d^4 x\,
               (v_{b,i} + v'_{b,i} )\, \Phi_i\, {\rm Tr}(F_b\wedge F_b)  , }
where $B_i$ is defined as the integral of the RR 5-form over the
corresponding 3-cycle and similarly $\Phi_i$ is defined
as the integral of the RR 3-form over the
corresponding 3-cycle.
Taking into account the Hodge duality between the fields $B_i$ and
$\Phi_{i+8}$ these axionic couplings indeed cancel the mixed anomalies.
For more details we refer the reader to the general discussion
in \rbbkl.

As was pointed out in \rimr\ the couplings \gsmech\
can generate a mass term  for $U(1)$ gauge fields
even if they are not anomalous.
The massless $U(1)$s are given by the kernel
of the matrix
\eqn\matric{   M_{ai}= N_a (v_{a,i} - v'_{a,i} )  .}
In our model it can be easily seen that $U(1)_2$ and $U(1)_3$
remain massless, so that the final gauge symmetry
is $SU(4)\times SU(2)\times SU(2)\times U(1)^2$.
We will not discuss this model any further but move forward
to the construction of a more realistic model with three generations.

\newsec{A 3 generation supersymmetric Pati-Salam model}

For the ${\bf ABB}$ model with the complex structure of the last
torus fixed at $U_2=1/2$, a computer search shows that only sufficiently small
mutual intersection numbers $(\pi_a\circ \pi_b,\pi'_a\circ
\pi_b)=(0,0),(\pm 1,0),(0,\pm 1)$ are possible. These numbers
allow the construction of a three generation model in the
following way. First, we start with seven stacks of D6-branes with
an initial gauge symmetry $U(4)\times U(2)^6$ and choose  the
wrapping numbers as shown in Table 6. 
\vskip 0.8cm 
\vbox{
\centerline{\vbox{ \hbox{\vbox{\offinterlineskip
\def\tablespace{height2pt&\omit&&\omit&&
 \omit&\cr}
\def\tablerule{\tablespace\noalign{\hrule}\tablespace}

\hrule\halign{&\vrule#&\strut\hskip0.2cm  #\hfill\hskip0.2cm\cr
& stack  && $(n_I,m_I)$  && homology cycle &\cr
\tablerule
& U(4)  && $(1,0;1,0;0,1)$ && $\pi_1={1\over 2}\left(
         \o\rho_1+\o\varepsilon_1+\o\varepsilon_3+\o\varepsilon_5\right)$ & \cr
& &&  && $\pi'_1={1\over 2}\left(
          \rho_2-\o\rho_2+\varepsilon_1-\varepsilon_3-\varepsilon_6-
             \o\varepsilon_1+\o\varepsilon_3+\o\varepsilon_6     \right)$ & \cr
\tablerule
\tablerule
& U(2)  && $(1,0;1,0;0,1)$ && $\pi_2={1\over 2}\left(
         \o\rho_1+\o\varepsilon_1-\o\varepsilon_3-\o\varepsilon_5\right)$ & \cr
& &&  && $\pi'_2={1\over 2}\left(
          \rho_2-\o\rho_2+\varepsilon_1+\varepsilon_3+\varepsilon_6-
             \o\varepsilon_1-\o\varepsilon_3-\o\varepsilon_6     \right)$ & \cr
\tablerule
& U(2)  && $(1,0;1,0;0,1)$ && $\pi_3={1\over 2}\left(
         \o\rho_1+\o\varepsilon_2-\o\varepsilon_3-\o\varepsilon_6\right)$ & \cr
& &&  && $\pi'_3={1\over 2}\left(
          \rho_2-\o\rho_2+\varepsilon_2+\varepsilon_3+\varepsilon_5-
             \o\varepsilon_2-\o\varepsilon_3-\o\varepsilon_5     \right)$ & \cr
\tablerule
& U(2)  && $(1,0;1,0;0,1)$ && $\pi_4={1\over 2}\left(
         \o\rho_1-\o\varepsilon_2-\o\varepsilon_3-\o\varepsilon_6\right)$ & \cr
& &&  && $\pi'_4={1\over 2}\left(
          \rho_2-\o\rho_2-\varepsilon_2+\varepsilon_3+\varepsilon_5+
             \o\varepsilon_2-\o\varepsilon_3-\o\varepsilon_5     \right)$ & \cr
\tablerule
\tablerule
& U(2)  && $(1,0;1,0;0,1)$ && $\pi_5={1\over 2}\left(
         \o\rho_1-\o\varepsilon_1+\o\varepsilon_3-\o\varepsilon_5\right)$ & \cr
& &&  && $\pi'_5={1\over 2}\left(
          \rho_2-\o\rho_2-\varepsilon_1-\varepsilon_3+\varepsilon_6+
             \o\varepsilon_1+\o\varepsilon_3-\o\varepsilon_6     \right)$ & \cr
\tablerule
& U(2)  && $(1,0;1,0;0,1)$ && $\pi_6={1\over 2}\left(
         \o\rho_1-\o\varepsilon_1-\o\varepsilon_4+\o\varepsilon_6\right)$ & \cr
& &&  && $\pi'_6={1\over 2}\left(
          \rho_2-\o\rho_2-\varepsilon_1+\varepsilon_4-\varepsilon_5+
             \o\varepsilon_1-\o\varepsilon_3+\o\varepsilon_5     \right)$ & \cr
\tablerule
& U(2)  && $(1,0;1,0;0,1)$ && $\pi_7={1\over 2}\left(
         \o\rho_1-\o\varepsilon_1+\o\varepsilon_4+\o\varepsilon_6\right)$ & \cr
& &&  && $\pi'_7={1\over 2}\left(
          \rho_2-\o\rho_2+\varepsilon_1-\varepsilon_4-\varepsilon_5+
             \o\varepsilon_1+\o\varepsilon_3+\o\varepsilon_5    \right)$ & \cr
}\hrule}}}} \centerline{ \hbox{{\bf Table 6:}{\it ~~ D6-branes for
3 generation PS-model}}} } 
\vskip 0.5cm 
\noindent 
Adding up all
homological 3-cycles, one realizes that the RR-tadpole
cancellation condition is satisfied. The contribution of the
O6-plane tension to the scalar potential is \eqn\tensa{
{V}_{O6}=-T_6\, e^{-\phi_4} 16 \sqrt{2}\left(
                    {1\over \sqrt{U_2}} +2 \sqrt{U_2} \right) ,}
whereas the seven stacks of D6-branes give
\eqn\tenso{\eqalign{
          {V}_{1}&=T_6\, e^{-\phi_4} 16 \sqrt{ {1\over 4\, U_2}+ U_2}    \cr
          {V}_{2,\ldots,7}&=T_6\, e^{-\phi_4} 8 \sqrt{{1\over 4\, U_2}+ U_2}.\cr}}
Adding up all terms, one finds that indeed the NS-NS tadpole
vanishes just for $U_2={1\over 2}$. This means that in contrast to
the four generation model, here supersymmetry really fixes the
complex structure of the third torus. This freezing of moduli for
supersymmetric backgrounds is very similar to what happens for
instance in recently discussed compactifications with
non-vanishing R-R fluxes \refs{\DasguptaSS,\rgkp,\rkst}.

In terms of ${\cal N}=2$ supermultiplets, the model contains
vector multiplets in the gauge group $U(4)\times U(2)^3\times
U(2)^3$ and in addition two hypermultiplets in the adjoint
representation of each unitary gauge factor. The complex scalar in
the vector multiplet corresponds to the unconstrained position of
each stack of D6-branes on the third $T^2$. As described in
section 4.2., the hypermultiplet appears on the intersection
between a stack of branes and its $\ZZ_4$ image. By computing the
intersection numbers, we derive the chiral spectrum as shown in
Table 7, where $n$ denotes the number of chiral multiplets in the
respective representation as given by the intersection number.
\vskip 0.8cm \vbox{ \centerline{\vbox{
\hbox{\vbox{\offinterlineskip
\def\tablespace{height2pt&\omit&&\omit&&
 \omit&\cr}
\def\tablerule{\tablespace\noalign{\hrule}\tablespace}

\hrule\halign{&\vrule#&\strut\hskip0.2cm \hfill #\hfill\hskip0.2cm\cr
& field && n  && $U(4)\times U(2)^3\times U(2)^3$   &\cr
\tablerule
& $\Phi_{1'2}$ && 1 && $(4;2,1,1;1,1,1)$ &\cr
& $\Phi_{1'3}$ && 1 && $(4;1,{2},1;1,1,1)$ &\cr
& $\Phi_{1'4}$ && 1 && $(4;1,1,{2};1,1,1)$ &\cr
\tablerule
& $\Phi_{1'5}$ && 1 && $(\o{4};1,1,1;\o{2},1,1)$ &\cr
& $\Phi_{1'6}$ && 1 && $(\o{4};1,1,1;1,\o{2},1)$ &\cr
& $\Phi_{1'7}$ && 1 && $(\o{4};1,1,1;1,1,\o{2})$ &\cr
\tablerule
& $\Phi_{2'3}$ && 1 && $(1;\o{2},\o{2},1;1,1,1)$ &\cr
& $\Phi_{2'4}$ && 1 && $(1;\o{2},1,\o{2};1,1,1)$ &\cr
& $\Phi_{3'4}$ && 1 && $(1;1,\o{2},\o{2};1,1,1)$ &\cr
\tablerule
& $\Phi_{5'6}$ && 1 && $(1;1,1,1;2,{2},1)$ &\cr
& $\Phi_{5'7}$ && 1 && $(1;1,1,1;2,1,{2})$ &\cr
& $\Phi_{6'7}$ && 1 && $(1;1,1,1;1,{2},{2})$ &\cr
}\hrule}}}}
\centerline{
\hbox{{\bf Table 7:}{\it ~~ Chiral spectrum for a 7-stack model}}}
}
\vskip 0.5cm
\noindent
First, we notice that all non-abelian anomalies cancel including
formally also the $U(2)$ anomalies.

In order to proceed and really get a three generation model, it is
necessary to break the two triplets $U(2)^3$ down to their
diagonal subgroups. Potential gauge symmetry breaking candidates
in this way are the chiral fields
$\{\Phi_{2'3},\Phi_{2'4},\Phi_{3'4}\}$ and
$\{\Phi_{5'6},\Phi_{5'7},\Phi_{6'7}\}$ from Table 7. However, one
has to remember that these are chiral ${\cal N}=1$ supermultiplets
living on the intersection of two D-branes in every case. Let us
review what massless bosons localized on intersecting D-branes
indicate.

\subsec{Brane recombination}

If two stacks of D-branes preserve a common ${\cal N}=2$ supersymmetry, then
a massless hypermultiplet, $H$, localized on the intersection,
signals a possible deformation of the two stacks of D-branes into 
recombined D-branes which wrap a complex cycle. Note, that two
factorizable branes can only preserve ${\cal N}=2$ supersymmetry
if they are parallel on one of the three $T^2_I$ tori.  The
complex cycle has the same volume as the sum of volumes of the two
D-branes before the recombination process occurs. In the effective
low energy theory, this recombination can be understood as a Higgs
effect where a flat direction $\langle h_1\rangle=\langle
h_2\rangle$ in the  D-term potential 
\eqn\dtermpo{  V_D={1\over 2
g^2}\left( h_1 \o{h}_1 - h_2 \o{h}_2\right)^2 } 
exists, along
which the $U(N)\times U(N)$ gauge symmetry is broken to the
diagonal subgroup \footnote{$^2$}
{If on one of the two stacks there sits only a single D6-brane, 
the F-term potential
$\phi h_1 h_2$ forbids the existence of a flat direction
with $\langle h_1\rangle=\langle
h_2\rangle$. This is the field theoretic correspondence  of the fact
that there do not exist large instantons in the $U(1)$ 
gauge group. We thank A. Uranga for pointing this out to us.}. 
Here $h_1$ and $h_2$ denote the two complex
bosons inside the hypermultiplet. Thus, in this case without
changing the closed string background, there exists an open string
modulus, which has the interpretation of a Higgs field in the low
energy effective theory. Note, that in the T-dual picture, this is
just the deformation of
a small instanton into an instanton of finite size.
In our concrete models such ${\cal N}=2$
Higgs sectors are coupled at brane intersections to chiral ${\cal
N}=1$ sectors. Note, that the brane recombination in the effective
gauge theory cannot simply be described by the renormalizable
couplings. In order to get the correct light spectrum, one also
has to take into account stringy higher dimensional couplings.

When the two D-branes only preserve ${\cal N}=1$ supersymmetry and
support a massless chiral supermultiplet $\Phi$ on the
intersection \refs{\rkachmca,\rwittena}, the situation gets a little bit
more involved. In this case, the analogous D-term potential is of
the form \eqn\dtermpoc{ V_D={1\over 2 g^2}\left( \phi  \o{\phi}
\right)^2 }
 which tells us that, unless there are more chiral fields involved, 
simply by giving a VEV to the massless boson
 $\phi$,
we do not obtain a flat direction of the D-term potential and
therefore break supersymmetry. Nevertheless, the massless modes
indicate that the intersecting brane configuration lies on a line
of marginal stability in the complex structure moduli space. By a
small variation of the complex structure, a Fayet-Iliopoulos (FI)
term, $r$, is introduced that changes the D-term potential  to
\eqn\dtermpoc{  V_D={1\over 2 g^2}\left( \phi \o{\phi} +r
\right)^2 .} Therefore, for $r<0$ the field $\phi$ becomes
tachyonic and there exists a new stable supersymmetric minimum of
the D-term potential. The intersecting branes then have combined
into one D-brane wrapping a special Lagrangian 3-cycle in the
underlying Calabi-Yau. For a finite FI-term $r$, this 3-cycle has
smaller volume than the two intersecting branes. However, the two
volumes are precisely equal on the line of marginal stability.
This means that on the line of marginal stability, there exists a
different configuration with only a single brane which also
preserves the same ${\cal N}=1$ supersymmetry and has the same
volume as the former pair of intersecting D-branes. Again the
gauge symmetry is broken to the diagonal subgroup. It has to be
emphasized that in this case the two configurations are not simply
linked by a Higgs mechanism in the effective low energy gauge
theory. As mentioned before, in order to deform the intersecting
brane configuration into the non-flat D-brane wrapping a special
Lagrangian 3-cycle, one first has to deform the closed string
background and then let the tachyonic mode condense. Therefore,
the description of this process is intrinsically stringy and
should be better described by string field theory rather than the
effective low  energy gauge theory\footnote{$^3$}{In the context
of so-called quasi-supersymmetric intersecting brane world models
\rqsusyb , it has been observed that indeed the brane
recombination of ${\cal N}=1$ supersymmetric intersections cannot
simply be described by a Higgs mechanism of massless modes. It was
suggested there that the stringy nature of this transition has the
meaning that also some massive, necessarily non-chiral, fields are
condensing during the brane recombination. At least from the
effective gauge theory point of view, this could induce the right
mass terms which are necessary for an understanding of the new
massless modes after the recombination. We leave it for future
work to find the right effective description of this transition,
but we can definitely state that it must involve some stringy
aspects as the complex structure changes, i.e. the closed string
background.}. For $r>0$, the non-supersymmetric intersecting
branes are stable and have a smaller volume than the recombined
brane. The lift of these brane recombination
processes to M-theory was discussed in \ruram.

After this little excursion, we come back to our model. We have
seen that the condensation of hypermultiplets is under much better
control than the condensation of chiral multiplets. Therefore, we
have to determine the Higgs fields in our model as well, meaning
to compute the non-chiral spectrum. This cannot be done by a
simple homology computation, but fortunately we do know the exact
conformal field theory at the orbifold point. Using the boundary
states \bound , we can determine the non-chiral matter living on
intersections of the various stacks of D-branes. One first
computes the overlap between two such boundary states and then
transforms the result to the open string channel to get the
annulus partition function, from which one can read off the
massless states. This is a straightforward but tedious
computation, which also confirms the chiral spectrum in Table 7.
Thus the conformal field theory result agrees completely with the
purely topological computation of the intersection numbers.

Computing the non-chiral spectrum just for one stack of $U(2)$
branes and their $\ZZ_4$ and $\Omega\o\sigma$ images, one first
finds the  well known hypermultiplet,
$\Phi_{adj}=(\phi_{adj},\tilde\phi_{adj})$, in the adjoint
representation of $U(2)$ localized on the intersection of a brane
and its $\ZZ_4$ image. Moreover, there are two chiral multiplets,
$\Psi_A$ and $\Psi_{\o A}$, in the ${\bf A}$ respectively ${\bf
\o{A}}$ representation arising from the $(\pi_i,\pi'_i)$ sector.
Since the two chiral fields carry conjugate representations of the
gauge group, they cannot be seen by the topological intersection
number which in fact vanishes, $\pi_i\circ\pi'_i=0$.
 We have depicted the resulting  quiver diagram for these three fields in
figure 5.
\fig{Adjoint higgsing}{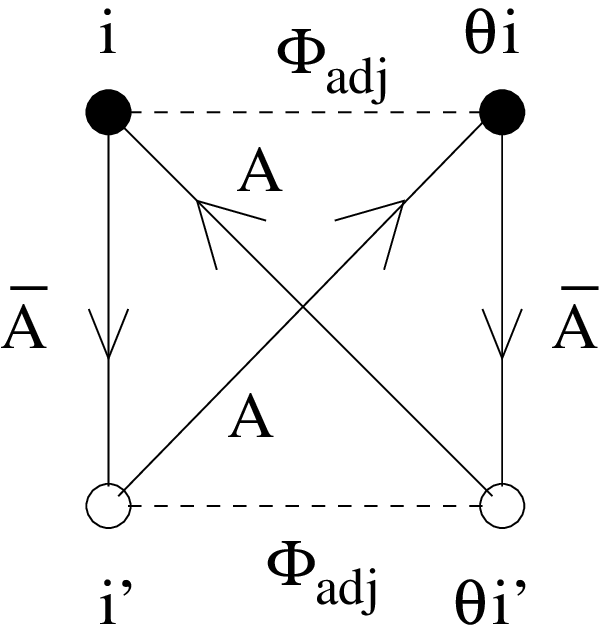}{6truecm}
\noindent
For each closed polygon in the quiver diagram, the
associated product of fields can occur in the holomorphic
superpotential. In our case, the following two terms can appear
\eqn\massterm{ W=\phi_{adj} \Psi_A\Psi_{\o A} + \tilde\phi_{adj}
\Psi_A\Psi_{\o A} ,}
which generate  a mass for  the
anti-symmetric fields when the adjoint multiplet gets a VEV.

As we have mentioned already in the last
section, giving a VEV to this adjoint field localized on the intersection
between a brane and its $\ZZ_4$ image, leads to the recombination
of these two branes. The recombined brane no longer passes through the
$\ZZ_4$ invariant intersection points.

After computing all annulus partition functions for
pairs of D-branes from Table 6, we find the total non-chiral spectrum
listed in Table 8.
\vskip 0.8cm
\vbox{
\centerline{\vbox{
\hbox{\vbox{\offinterlineskip
\def\tablespace{height2pt&\omit&&\omit&&
 \omit&\cr}
\def\tablerule{\tablespace\noalign{\hrule}\tablespace}

\hrule\halign{&\vrule#&\strut\hskip0.2cm \hfill #\hfill\hskip0.2cm\cr
& field &&  n  && $U(4)\times U(2)^3\times
U(2)^3$   &\cr \tablerule & $H_{12}$ && 1 &&
$(4;\o{2},1,1;1,1,1)+c.c.$ &\cr
& $H_{13}$ && 1 && $(4;1,\o{2},1;1,1,1)+c.c.$ &\cr
& $H_{14}$ && 1 && $(4;1,1,\o{2};1,1,1)+c.c.$ &\cr
\tablerule
& $H_{15}$ && 1 && $(\o{4};1,1,1;{2},1,1)+c.c.$ &\cr
& $H_{16}$ && 1 && $(\o{4};1,1,1;1,{2},1)+c.c.$ &\cr
& $H_{17}$ && 1 && $(\o{4};1,1,1;1,1,{2})+c.c.$ &\cr
\tablerule
& $H_{25}$ && 1 && $(1;{2},1,1;\o{2},1,1)+c.c.$ &\cr
& $H_{26}$ && 1 && $(1;{2},1,1;1,\o{2},1)+c.c.$ &\cr
& $H_{27}$ && 1 && $(1;{2},1,1;1,1,\o{2})+c.c.$ &\cr
& $H_{35}$ && 1 && $(1;1,{2},1;\o{2},1,1)+c.c.$ &\cr
& $H_{36}$ &&  1 && $(1;1,{2},1;1,\o{2},1)+c.c.$ &\cr
& $H_{37}$ && 1 && $(1;1,{2},1;1,1,\o{2})+c.c.$ &\cr
& $H_{45}$ && 1 && $(1;1,1,{2};\o{2},1,1)+c.c.$ &\cr
& $H_{46}$ && 1 && $(1;1,1,{2};1,\o{2},1)+c.c.$ &\cr
& $H_{47}$ && 1 && $(1;1,1,{2};1,1,\o{2})+c.c.$ &\cr
}\hrule}}}}
\centerline{
\hbox{{\bf Table 8:}{\it ~~ Non-chiral spectrum (Higgs fields)}}}
}
\vskip 0.5cm
\noindent
It is interesting that we find  Higgs fields  which might break a
$SU(4)\times SU(2)\times SU(2)$ gauge symmetry in a first step
down to the  Standard Model and in a second step down to
$SU(3)\times U(1)_{em}$. However, the Higgs fields which would
allow us to break the product groups $U(2)^3$ down to their
diagonal subgroup are not present in the non-chiral spectrum.

\subsec{D-flatness}

However, we do have the massless chiral bifundamental fields
$\{\Phi_{2'3},\ldots,\Phi_{6'7}\}$ living on  intersections
preserving ${\cal N}=1$ supersymmetry.
As we have already mentioned, for isolated brane intersections
these massless fields indicate
that the complex structure moduli are chosen such that
one sits on a line of marginal stability. On one side
of this line, the intersecting branes break supersymmetry
without developing a tachyonic mode. This indicates that
the intersecting brane configuration is stable.
But on the other side of the line, the former massless
chiral field becomes tachyonic and after condensation
leads to a new in general non-flat supersymmetric brane
wrapping a special Lagrangian 3-cycle. Since the tachyon
transforms in the bifundamental representation, on this brane
the gauge symmetry is broken to its diagonal subgroup.


We therefore expect for our compact situation that at least
locally these bifundamental chiral multiplets indicate the
existence of a recombined brane of the same volume but with the
gauge group broken to the diagonal subgroup. In order to make our
argument save, we need to show that the D-terms allow, that for
certain continuous deformations of the complex structure moduli,
just the four fields
$\{\Phi_{2'3},\Phi_{2'4},\Phi_{5'6},\Phi_{5'7}\}$  become
tachyonic. Then they condense to a new supersymmetric ground state
and the gauge symmetries $U(2)^3$ are broken to the diagonal
$U(2)$s. From general arguments for open string models
with ${\cal N}=1$ supersymmetry, it is known that the complex structure moduli only
appear in the D-term potential, whereas the K\"ahler moduli only
appear in the F-term potential \refs{\rbdlr,\rkklma\rkklmb,\rav}.

Remember that the Green-Schwarz mechanism requires the Chern-Simons
couplings to be of the form
\eqn\gsmechb{  S_{CS}=   \sum_{i=1}^{b_3} \sum_{a=1}^k \int d^4 x\, M_{ai}\, B_i\wedge
               {1\over N_a}\,{\rm tr}(F_a) .  }
The supersymmetric completion involves a coupling of the
auxiliary field $D_a$
\eqn\fib{  S_{FI}=  \sum_{i=1}^{b_3} \sum_{a=1}^k \int d^4 x\,  M_{ai}\,
{\partial{\cal K}\over \partial \phi_i} \,
               {1\over N_a}\,{\rm tr}(D_a) ,  }
where $\phi_i$ are the superpartners of the Hodge duals of
the RR 2-forms and ${\cal K}$ denotes the K\"ahler potential.
Thus, these couplings give rise to FI-terms depending
on the complex structure moduli which we parameterize
simply by $A_i={\partial{\cal K}/\partial \phi_i}$
\footnote{$^4$}{For our purposes we do not need the precise form of
the K\"ahler potential as long as the map from the complex structure
moduli $\phi_i$ to the new parameters $A_i$ is one to one. But this
is the case, as the functional determinant for the map between
these two sets of variables is equal to
Det$\left({\partial^2 {\cal K}\over \partial \phi_i\partial \phi_j}\right)$,
which is non-vanishing for a positive definite metric
on the complex structure moduli space.}.

Let us now discuss the D-term potential for the $U(4)\times U(2)^3\times
U(2)^3$ gauge fields
in our model and see whether it allows supersymmetric
ground states of the type described above.
The D-term potential including only the chiral matter and
the FI-terms  in general reads
\eqn\dtermp{\eqalign{   V_D&=\sum_{a=1}^k
 \sum_{r,s=1}^{N_a} {1\over 2 g_a^2}(D^{rs}_a)^2 \cr
                 &=\sum_{a=1}^k \sum_{r,s=1}^{N_a}
       {1\over 2 g_a^2}\left(\sum_{j=1}^k\sum_{p=1}^{N_j} q_{aj}
       \, \Phi^{rp}_{aj} \,
          \o\Phi^{sp}_{aj}+
            g_a^2\sum_{i=1}^{b_3}  {M_{ai}\over N_a}\, A_i\,
                 \delta^{rs}\right)^2 ,\cr}}
where the indices $(r,s)$ numerate the $N_a^2$ gauge fields in the adjoint
representation
of the gauge factor $U(N_a)$ and the sum over $j$ is over all
chiral fields charged under $U(N_a)$.
The gauge coupling constants depend on the complex
structure moduli as well, but since we are only interested in the leading order
effects, we can set them to the constant values on the line of marginal
stability.  Since all branes have the same volume there, in the following we
will simply set them to one.
In our case, the charges $q_{aj}$ can be read off from Table 7 and
the Green-Schwarz  couplings $M_{ai}$ from Table 6 using the definition \matric.
It is then straightforward  to show that for the following non-vanishing
$\Omega\o\sigma$ invariant complex structure deformations related to the
four 3-cycles $\{\varepsilon_1,\varepsilon_2,\varepsilon_3-2\o\varepsilon_3,
\varepsilon_4-2\o\varepsilon_4\}$
\eqn\compd{\eqalign{
                A_{3}&=-{\kappa},\quad\quad \phantom{l2\lambda}
                   A_{5}-2A_{13}=-{\kappa}  \cr
                A_{4}&={\kappa}-2\lambda,\quad\quad
                A_{6}-2A_{14}=2\mu-{\kappa} \cr}}
just the fields
$\{\Phi_{2'3},\Phi_{2'4},\Phi_{5'6},\Phi_{5'7}\}$
become tachyonic. There exists a new supersymmetric
ground state for the non-vanishing VEVs
\eqn\compd{\eqalign{
                |\Phi^{rr}_{2'3}|^2&={\lambda},\quad\quad
                |\Phi^{rr}_{2'4}|^2={\kappa-\lambda} \cr
                |\Phi^{rr}_{5'6}|^2&={\mu},\quad\quad
                |\Phi^{rr}_{5'7}|^2={\kappa-\mu} \cr}}
with $r=1,2$, $\lambda,\mu>0$, $\kappa>\lambda$ and $\kappa>\mu$.
From this small calculation, we conclude that our model
indeed sits on a locus  of marginal stability, for which  a
supersymmetric configuration exists where
the branes $\{\pi_2,\pi'_3,\pi'_4\}$ and similarly the branes
$\{\pi_5,\pi'_6,\pi'_7\}$ have recombined into a single
stack of branes within the same homology class.

\subsec{Gauge symmetry breaking}

After this recombination process we are left with only three stacks of
D6-branes wrapping the homology cycles
\eqn\recom{   \pi_a=\pi_1, \quad \pi_b=\pi_2+\pi'_3+\pi'_4,
              \quad \pi_c=\pi_5+\pi'_6+\pi'_7 .} These branes are
not factorizable but we have presented arguments ensuring
that they preserve the same supersymmetry as the closed string
sector and the former intersecting brane configuration
\footnote{$^5$}{Since we get chiral fields in the (anti-)symmetric
representations after brane recombination, one might check
if those intersection numbers can also be obtained by flat
factorizable D-branes. Remember that we had the first assumption
that there are no such chiral fields in the (anti-)symmetric
representations. In fact, after an extensive computer search we
have not been able to find a model with just factorizable D-branes
generating the chiral spectrum of Table 9.}. The chiral spectrum
for this now 3 stack model is shown in Table 9.
\vskip 0.8cm
\vbox{
\centerline{\vbox{ \hbox{\vbox{\offinterlineskip
\def\tablespace{height2pt&\omit&&\omit&&
 \omit&\cr}
\def\tablerule{\tablespace\noalign{\hrule}\tablespace}

\hrule\halign{&\vrule#&\strut\hskip0.2cm \hfill #\hfill\hskip0.2cm\cr
& field && n  && $SU(4)\times SU(2)\times SU(2)\times U(1)^3$   &\cr
\tablerule
& $\Phi_{ab}$  && 2 && $(4,2,1)_{(1,-1,0)}$ &\cr
& $\Phi_{a'b}$  &&  1 && $(4,2,1)_{(1,1,0)}$ &\cr
\tablerule
& $\Phi_{ac}$  && 2 && $(\o{4},1,2)_{(-1,0,1)}$ &\cr
& $\Phi_{a'c}$  &&  1 && $(\o{4},1,2)_{(-1,0,-1)}$ &\cr
\tablerule
& $\Phi_{b'b}$  && 1 && $(1,S+A,1)_{(0,2,0)}$ &\cr
& $\Phi_{c'c}$  &&  1 && $(1,1,\o{S}+\o{A})_{(0,0,-2)}$ &\cr
}\hrule}}}}
\centerline{
\hbox{{\bf Table 9:}{\it ~~ Chiral spectrum for 3 stack PS-model}}}
}
\vskip 0.5cm
\noindent
The intersection numbers $\pi'_{b,c}\circ \pi_{b,c}$ do
not vanish any longer, therefore giving  rise to chiral
multiplets in the symmetric and anti-symmetric representation of the $U(2)$
gauge factors. Clearly, these chiral fields are needed in order to cancel the
formal non-abelian $U(2)$ anomalies. These anti-symmetric
fields can be understood as the remnants of the chiral fields, $\Phi_{3'4}$
and $\Phi_{6'7}$,
which did not condense during the brane recombination process.

Computing the mixed anomalies for this model, one finds that
two $U(1)$ gauge factors are anomalous and that the  only anomaly free
combination is
\eqn\freean{    U(1)=U(1)_a-3\, U(1)_b -3\, U(1)_c .}
The quadratic axionic couplings reveal
that the matrix $M_{ai}$ in \matric\
has a trivial kernel and therefore
all three $U(1)$ gauge groups become massive and survive
as global symmetries.
To summarize, after the recombination of some of the $U2)$ branes
we have found  a supersymmetric 3 generation
Pati-Salam model with gauge group $SU(4)\times SU(2)_L\times SU(2)_R$
which accommodates the standard model matter in addition
to some exotic matter in the (anti-)symmetric representation
of the $SU(2)$ gauge groups.

To compute the massless non-chiral spectrum after the recombination,
we have to determine which Higgs fields receive a mass from  couplings
with the condensing chiral bifundamental fields.
As we have explained earlier, the applicability of the low energy
effective field theory is limited but still is the only
information we have. So, we will see how far we can get.
We first consider the sector of the branes $\{\pi_1, \ldots, \pi_4\}$ in
figure 6.
\fig{Quiver diagram for the branes $\{1,2,3,4\}$}{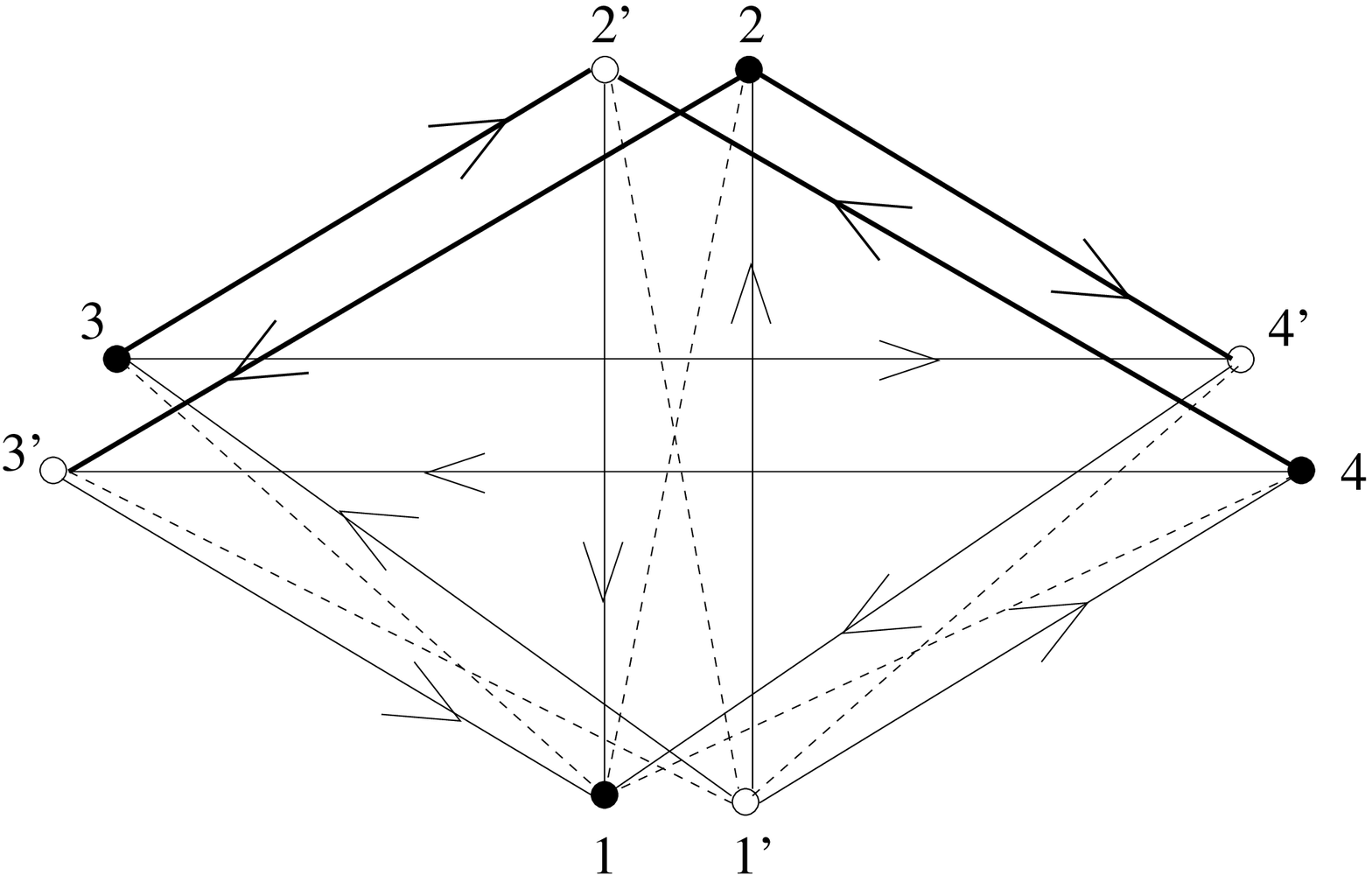}{14truecm}
\noindent
The chiral fields are indicated by an arrow and non-chiral fields
by a dashed line. The fields which receive a VEV after small complex
structure deformations are depicted by a fat line.
Let us decompose  the Higgs fields inside one hypermultiplet into its
two chiral components $H_{1j}=(h^{(1)}_{1j}, h^{(2)}_{1j})$ for $j=2,3,4$.
We observe a couple of closed  triangles in the quiver diagram in figure 6
that give rise to the following Yukawa couplings in the superpotential
\eqn\supercoup{\eqalign{    &\Phi_{2'3}\, \Phi_{1'2}, h^{(2)}_{13},\quad\quad
                            \Phi_{2'3}\, \Phi_{1'3}, h^{(2)}_{12} \cr
                            &\Phi_{2'4}\, \Phi_{1'2}, h^{(2)}_{14},\quad\quad
                            \Phi_{2'4}\, \Phi_{1'4}, h^{(2)}_{12}. \cr}}
Condensation of the chiral fields $\Phi_{2'3}$ and $\Phi_{2'4}$
leads to a mass matrix for the six fields $\{\Phi_{1'2},
\Phi_{1'3}, \Phi_{1'4},h^{(2)}_{12},h^{(2)}_{13}, h^{(2)}_{14}\}$
of rank four. Thus, one combination of the three fields $\Phi$,
one combination of the three fields $h^{(2)}$ and furthermore the three fields
$h^{(1)}$ remain massless. These modes just fit into the three
chiral fields in Table 9 in addition to one further hypermultiplet
in the $(4,{2},1)$ representation of the Pati-Salam gauge group
$U(4)\times U(2)\times U(2)$. The condensation for the second
triplet of $U(2)$s is completely analogous and leads to a
massless hypermultiplet in the $(4,1,{2})$ representation.
\noindent
\fig{Quiver diagram for the branes
$\{2,3,4,5,6,7\}$}{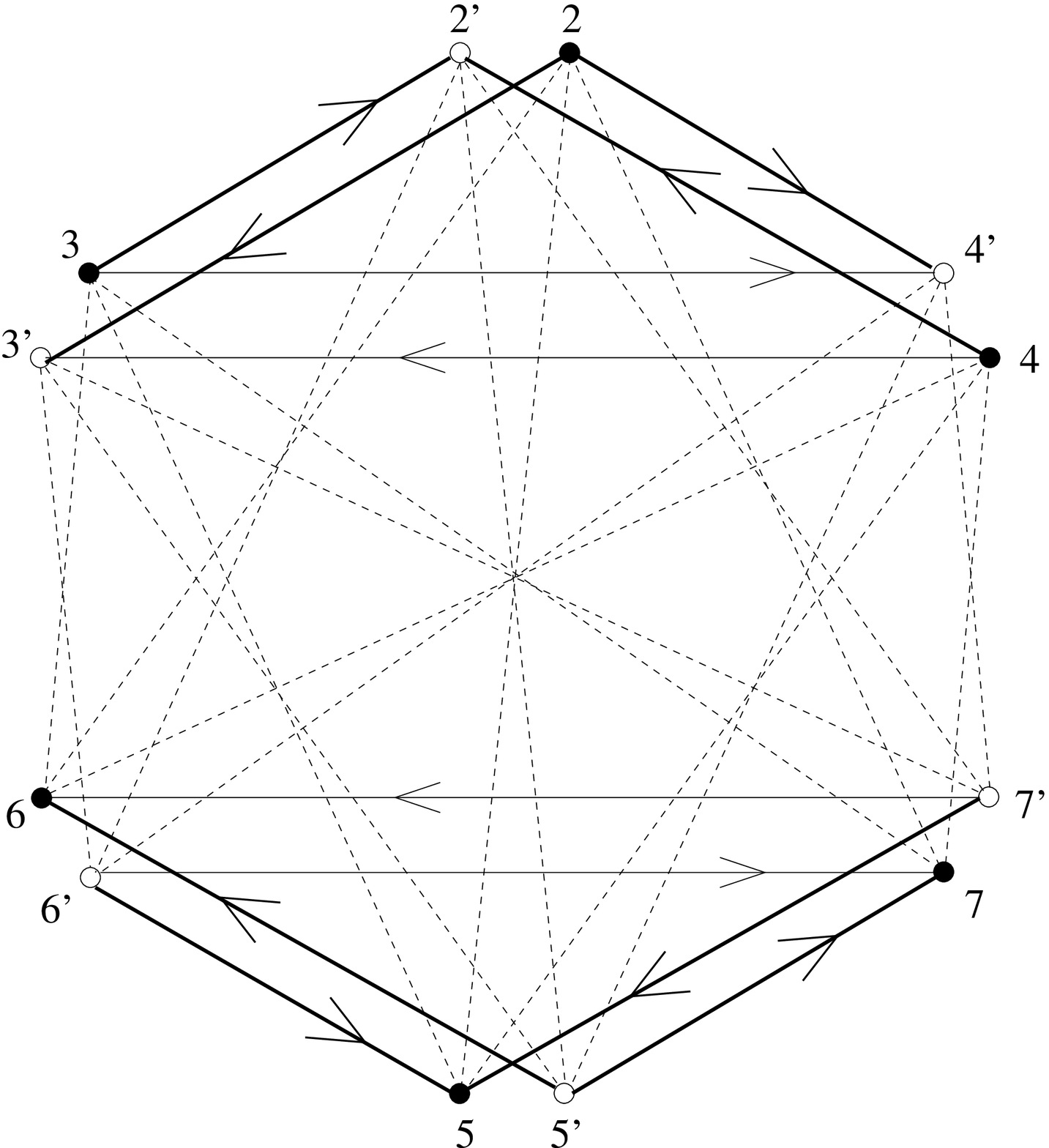}{16truecm}
\noindent
The quiver
diagram involving the six $U(2)$ gauge groups is shown in figure
7. In this quiver diagram there are closed polygons like
$(2-4'-7'-6)$ which after condensation generate a mass term for
one chiral component inside each of the nine hypermultiplets
$\{H_{25},H_{26},\ldots,H_{46},H_{47}\}$. Remember that a
hypermultiplet consists of two chiral multiplets of opposite
charge, $H=(h^{(1)},h^{(2)})$. The mass matrix for these nine
chiral fields has rank six, so that three combinations of the four
chiral fields, $h^{(1)}$, in $\{H_{36},H_{37},H_{46},H_{47}\}$
remain massless. Since the intersection numbers in Table 9 tell us
that there are no chiral fields in the $(1,2,2)$ representation of
the $U(4)\times U(2)\times U(2)$ gauge group, the other chiral
components, $h^{(2)}$, of the hypermultiplets must also gain a mass
during brane recombination. A very similar behavior was found in \rqsusyb,
and it was pointed out that this might involve the
condensation of massive string modes, as well. These
would at least allow the correct mass terms in the quiver diagram.

We expect that the quiver diagram really tells us half of the complete
story, so that the non-chiral spectrum of the
three generation Pati-Salam model is as listed in Table 10.
\vskip 0.8cm
\vbox{
\centerline{\vbox{
\hbox{\vbox{\offinterlineskip
\def\tablespace{height2pt&\omit&&\omit&&
 \omit&\cr}
\def\tablerule{\tablespace\noalign{\hrule}\tablespace}

\hrule\halign{&\vrule#&\strut\hskip0.2cm \hfill #\hfill\hskip0.2cm\cr
& field && n  && $U(4)\times U(2)\times U(2)$   &\cr
\tablerule
& $H_{aa}$  && 1 && $({\rm Adj},1,1)+c.c.$ &\cr
& $H_{bb}$  && 1 && $(1,{\rm Adj},1)+c.c.$ &\cr
& $H_{cc}$  && 1 && $(1,1,{\rm Adj})+c.c.$ &\cr
\tablerule
& $H_{a'b}$  && 1 && $(4,{2},1)+c.c.$ &\cr
\tablerule
& $H_{a'c}$  && 1 && $(4,1,{2})+c.c.$ &\cr
\tablerule
& $H_{bc}$  && 3 && $(1,2,\o{2})+c.c.$ &\cr
}\hrule}}}}
\centerline{
\hbox{{\bf Table 10:}{\it ~~ Non-chiral spectrum for 3 stack PS-model}}}
}
\noindent
Intriguingly, these are just appropriate Higgs fields to break
the Pati-Salam gauge group down to the Standard Model.

\subsec{Getting the Standard Model}

It is beyond the scope of this paper to discuss all the
phenomenological consequences of this 3 generation Pati-Salam model.
However, we would like to present two possible ways of breaking
the GUT Pati-Salam model down to
the Standard Model.
\vfill\eject
\meno
{\it 6.4.1. Adjoint Pati-Salam breaking}
\smno

There are still the adjoint scalars related
to the unconstrained positions of the branes on the third
$T^2$. By moving one of the four D6-branes away from the $U(4)$ stack,
or in other words by giving VEV to appropriate fields in the adjoint
of $U(4)$, we can break the gauge group down to
$U(3)\times U(2)\times U(2)\times U(1)$. Indeed the resulting spectrum
as shown in Table 11  looks like a three generation left-right
symmetric extension of the standard model.
\vskip 0.2cm
\vbox{
\centerline{\vbox{
\hbox{\vbox{\offinterlineskip
\def\tablespace{height2pt&\omit&&\omit&&
 \omit&\cr}
\def\tablerule{\tablespace\noalign{\hrule}\tablespace}

\hrule\halign{&\vrule#&\strut\hskip0.2cm \hskip 0.3cm #\hfill\hskip0.2cm\cr
& n  && $SU(3)_c\times SU(2)_L\times SU(2)_R\times U(1)^4$  &&
             $U(1)_{B-L}$ &\cr
\tablerule
& 1 && $(3,2,1)_{(1,1,0,0)}$ && ${1\over 3}$ &\cr
& 2 && $(3,2,1)_{(1,-1,0,0)}$ && ${1\over 3}$&\cr
\tablerule
& 1 && $(\o{3},1,2)_{(-1,0,-1,0)}$ && $-{1\over 3}$ &\cr
& 2 && $(\o{3},1,2)_{(-1,0,1,0)}$ && $-{1\over 3}$ &\cr
\tablerule
& 1 && $(1,2,1)_{(0,1,0,1)}$ && ${-1}$ &\cr
& 2 && $(1,2,1)_{(0,-1,0,1)}$ && ${-1}$ &\cr
\tablerule
& 1 && $(1,1,2)_{(0,0,-1,-1)}$ && ${1}$ &\cr
& 2 && $(1,1,2)_{(0,0,1,-1)}$ && ${1}$ &\cr
\tablerule
& 1 && $(1,S+A,1)_{(0,2,0,0)}$ && $0$ &\cr
& 1 && $(1,1,\o{S}+\o{A})_{(0,0,-2,0)}$ && $0$ &\cr
}\hrule}}}}
\centerline{
\hbox{{\bf Table 11:}{\it ~~ Chiral spectrum for 4 stack left-right symmetric SM}}}
}
\vskip 0.5cm
\noindent
Performing the anomaly analysis, one finds
two anomaly free $U(1)$s, of which
the combination ${1\over 3}(U(1)_1-3U(1)_4)$ remains massless
even after the Green-Schwarz mechanism. This linear combination
in fact is the $U(1)_{B-L}$ symmetry, which is expected to be anomaly-free
in a model with right-handed neutrinos.

By giving a VEV to fields
in the adjoint of $U(2)_R$, one obtains the next symmetry breaking, where the two $U(2)_R$
branes split into two $U(1)$ branes. This gives rise to the gauge symmetry
$U(3)\times U(2)_L\times U(1)_R\times U(1)_R\times U(1)$.
In this case the following two $U(1)$ gauge factors remain
massless after checking the
Green-Schwarz couplings
\eqn\massu{\eqalign{    U(1)_{B-L}&={1\over 3}(U(1)_1-3U(1)_5) \cr
                        U(1)_Y&={1\over 3}U(1)_1 +
U(1)_3-U(1)_4-U(1)_5 .\cr}}
It is very assuring that we indeed
obtain a massless hypercharge. The final supersymmetric chiral
spectrum is listed in Table 12 with respect to the unbroken gauge
symmetries.
\vskip 0.8cm
\vbox{ \centerline{\vbox{
\hbox{\vbox{\offinterlineskip
\def\tablespace{height2pt&\omit&&\omit&&\omit&&
 \omit&\cr}
\def\tablerule{\tablespace\noalign{\hrule}\tablespace}

\hrule\halign{&\vrule#&\strut\hskip0.2cm \hskip 0.3cm #\hfill\hskip0.2cm\cr
& n  && field && $SU(3)\times SU(2)\times U(1)^3$ && $U(1)_Y \times U(1)_{B-L}$  &\cr
\tablerule
& 1 &&  $q_L$ && $(3,2)_{(1,1,0,0,0)}$ &&  $\left({1\over 3},{1\over
    3}\right)$ &\cr
& 2 &&  $q_L$ && $(3,2)_{(1,-1,0,0,0)}$ &&  $\left({1\over 3},{1\over
    3}\right)$ &\cr
\tablerule
& 1 &&  $u_R$ && $(\o{3},1)_{(-1,0,-1,0,0)}$ &&  $\left(-{4\over 3},-{1\over
    3}\right)$ &\cr
& 2 &&  $u_R$ && $(\o{3},1)_{(-1,0,0,1,0)}$ &&  $\left(-{4\over 3},-{1\over
    3}\right)$ &\cr
& 2 &&  $d_R$ && $(\o{3},1)_{(-1,0,1,0,0)}$ &&  $\left({2\over 3},-{1\over
    3}\right)$ &\cr
& 1 &&  $d_R$ && $(\o{3},1)_{(-1,0,0,-1,0)}$ &&  $\left({2\over 3},-{1\over
    3}\right)$ &\cr
\tablerule
& 1 &&  $l_L$ && $(1,2)_{(0,1,0,0,1)}$ &&  $\left(-{1},-{1}\right)$ &\cr
& 2 &&  $l_L$ && $(1,2)_{(0,-1,0,0,1)}$ &&  $\left(-{1},-{1}\right)$ &\cr
\tablerule
& 2 &&  $e_R$ && $(1,1)_{(0,0,1,0,-1)}$ &&  $\left({2},{1}\right)$ &\cr
& 1 &&  $e_R$ && $(1,1)_{(0,0,0,-1,-1)}$ &&  $\left({2},{1}\right)$ &\cr
& 1 &&  $\nu_R$ && $(1,1)_{(0,0,-1,0,-1)}$ &&  $\left({0},{1}\right)$ &\cr
& 2 &&  $\nu_R$ && $(1,1)_{(0,0,0,1,-1)}$ &&  $\left({0},{1}\right)$ &\cr
\tablerule
& 1 &&  $$ && $(1,S+A)_{(0,2,0,0,0)}$ && $\left({0},{0}\right)$  &\cr
& 1 &&  $$ && $(1,1)_{(0,0,-2,0,0)}$ && $\left(-{2},{0}\right)$  &\cr
& 1 &&  $$ && $(1,1)_{(0,0,0,-2,0)}$ && $\left({2},{0}\right)$  &\cr
& 2 &&  $$ && $(1,1)_{(0,0,-1,-1,0)}$ && $\left({0},{0}\right)$  &\cr
}\hrule}}}}
\centerline{
\hbox{{\bf Table 12:}{\it ~~ Chiral spectrum for 5 stack SM}}}
}
\vskip 0.5cm
\noindent
The anomalous $U(1)_1$ can be identified with the baryon number
operator and survives the Green-Schwarz mechanism as a global
symmetry. Therefore, in this model the baryon number
is conserved and the proton is stable.
Similarly, $U(1)_5$ can be identified with the lepton number
and also survives as a global symmetry.
To break the gauge symmetry $U(1)_{B-L}$, one can recombine
the third and the fifth stack of D6 branes, which is expected to
 correspond  to giving
a VEV to the Higgs field $H_{3'5}$. We will see in section 6.4.2. that
this brane recombination gives a mass to the right handed neutrino.

To proceed, let us compute the relation between the Standard Model
gauge couplings at the PS-breaking   scale at string tree
level. The $U(N_a)$ gauge couplings for D6-branes are given by
\eqn\gaugec{ {4\pi\over g_a^2}={M^3_s\over g_s } {\rm Vol(D6_a)} ,}
where Vol$(D6_a)$ denotes the internal volume of the 3-cycle the
D6-branes are  wrapping on. During the brane recombination process the
volume of the recombined brane is equal to the sum of the volumes
of the two intersecting branes. Therefore, we have the
following ratios for the volumes of the five stacks of D6-branes in our model
\eqn\ratio{ {\rm Vol(D6_2)}={\rm Vol(D6_3)}={\rm Vol(D6_4)}=3{\rm
Vol(D6_1)}, \quad
             {\rm Vol(D6_5)}={\rm Vol(D6_1)} .}
This allows us at string tree level to determine the ratio of the Standard Model
gauge couplings at the PS breaking scale to be
\eqn\ratiob{    {\alpha_{s}\over \alpha_Y}={11\over 3}, \quad\quad
                {\alpha_{w}\over \alpha_Y}={11\over 9} }
leading to a Weinberg angle $\sin^2 \theta_w=9/20$ which differs
from the usual $SU(5)$ GUT prediction
$\sin^2(\theta_w)=3/8$. Encouragingly, from \ratiob\ we get the
right order for the sizes of the Standard Model gauge couplings
constants, $\alpha_{s}> \alpha_{w}>\alpha_Y$.
It would be interesting to analyze
whether this GUT value is
consistent with the low energy data at the weak scale. A potential
problem is the  appearance of colored Higgs fields in Table 10,
which would spoil the asymptotic freedom of the $SU(3)$. In  order to
improve this situation one needs a model with less
non-chiral matter, i.e. a model where not so many open string sectors
actually preserve ${\cal N}=2$ supersymmetry.

\bigno {\it 6.4.2. Bifundamental  Pati-Salam breaking}
\meno
We can also use directly the bifundamental Higgs fields like
$H_{a'c}$ to break the model down to the Standard Model gauge
group. This higgsing in string theory should correspond to a
recombination of one of the four D6-branes wrapping $\pi_a$ with one of
the branes wrapping $\pi'_c$. Thus, we get the following four
stacks of D6-branes
\eqn\recomp{\eqalign{ \pi_A=\pi_a, \quad
\pi_B=\pi_b, \quad \pi_C=\pi_a+\pi'_c, \quad \pi_D=\pi_c}}
supporting  the initial gauge group $U(3)\times U(2)\times
U(1)^2$. The tadpole cancellation conditions are still satisfied.
One gets the chiral spectrum by computing the homological
intersection numbers as shown in Table 13.
\vskip 0.8cm \vbox{
\centerline{\vbox{ \hbox{\vbox{\offinterlineskip
\def\tablespace{height2pt&\omit&&\omit&&\omit&&\omit&&
 \omit&\cr}
\def\tablerule{\tablespace\noalign{\hrule}\tablespace}

\hrule\halign{&\vrule#&\strut\hskip0.2cm \hskip 0.3cm #\hfill\hskip0.2cm\cr
& n && field  && sector && $SU(3)_c\times SU(2)_L\times U(1)^4$ && $U(1)_Y$ &\cr
\tablerule
& 2 &&   $q_L$ &&  $(AB)$ && $(3,2)_{(1,-1,0,0)}$  && ${1\over 3}$ &\cr
& 1 && $q_L$ &&  $(A'B)$ && $(3,2)_{(1,1,0,0)}$  && ${1\over 3}$ &\cr
\tablerule
& 1 && $u_R$ &&  $(AC)$ && $(\o{3},1)_{(-1,0,1,0)}$  && $-{4\over 3}$ &\cr
& 2 && $d_R$ &&  $(A'C)$ && $(\o{3},1)_{(-1,0,-1,0)}$   && ${2\over 3}$  &\cr
\tablerule
& 2 && $u_R$ &&  $(AD)$ && $(\o{3},1)_{(-1,0,0,1)}$  && $-{4\over 3}$ &\cr
& 1 && $d_R$ &&  $(A'D)$ && $(\o{3},1)_{(-1,0,0,-1)}$ && ${2\over 3}$  &\cr
\tablerule
& 2 && $l_L$ &&  $(BC)$ && $(1,2)_{(0,-1,1,0)}$  && $-{1}$  &\cr
& 1 && $l_L$ &&  $(B'C)$ && $(1,2)_{(0,1,1,0)}$  && $-{1}$  &\cr
\tablerule
& 1 && $e_R$ &&  $(C'D)$ && $(1,1)_{(0,0,-1,-1)}$   && ${2}$ &\cr
& 1 && $e_R$ && $(C'C)$   && $(1,1)_{(0,0,-2,0)}$ && $2$ &\cr
& 1 && $e_R$ && $(D'D)$   && $(1,1)_{(0,0,0,-2)}$ && $2 $ &\cr
\tablerule
& 1 && $S$  && $(B'B)$   && $(1,S+A)_{(0,2,0,0)}$ && $0$ &\cr
}\hrule}}}}
\centerline{
\hbox{{\bf Table 13:}{\it ~~ Chiral spectrum for 4 stack SM}}}
}
\vskip 0.5cm
\noindent
By computing the mixed anomalies, one finds that there are
two anomalous $U(1)$ gauge factors and two anomaly free ones
\eqn\massu{\eqalign{    U(1)_Y&={1\over 3}U(1)_A-U(1)_C - U(1)_D \cr
                        U(1)_{K}&=U(1)_A-9\, U(1)_B +9\, U(1)_C-9\,
U(1)_D. \cr }}
Remarkably, the axionic couplings just leave the hypercharge massless,
so that we finally get the Standard Model gauge group
$SU(3)_C\times SU(2)_L\times U(1)_Y$.
In this model only the baryon number generator can be identified with $U(1)_1$, whereas
the lepton number is broken. Therefore, the proton is stable and lepton number
violating couplings as Majorana mass terms are possible.
Note, that there are no  massless right-handed neutrinos
in this model. As we have mentioned already, this model is related
to the model discussed in the last section by a further brane recombination
process, affecting the mass of the right handed neutrinos.
This brane recombination can be considered as a stringy mechanism
to generate GUT scale masses for the right handed neutrinos \rqsusyb.
The different ways of gauge symmetry breaking that have been discussed so far are depicted
in figure 8.
\smno
\begingroup
\midinsert
\global\advance\figno by 1
\centerline{
\xymatrix@=7pt@R=14pt{
         \save-<1.7em,1em>
         *\txt{ $1^{\rm st}$ adjoint \\ breaking} \restore &
            & U(4)\ar[dl]\ar[dr] &  & \times
            & & U(2)_R \ar[d]  & & \times & U(2)_L\ar[d]       \\
                  \save-<1.7em,1.2em>
         *\txt{ $2^{\rm nd}$ adjoint \\ breaking} \restore
          & U(3)\ar[d] & \times & U(1)\ar[d] & \times
                            & & U(2)_R\ar[dl]\ar[dr]&  & \times
           & U(2)_L\ar[d]   \\
           & U(3)\ar[d] & \times &
           U(1)\ar[dr]
                   & \times
                   & U(1)_R\ar[dl] & \times &  U(1)_R\ar[d]
                   & \times & U(2)_L\ar[d] \\
         \save-<1.7em,-1.5em>
         *\txt{ bifund. \\ breaking} \restore
           & U(3) & \times & &  U(1) & \ar@{=>}[d] & \times & U(1)_R
          & \times & U(2)_L
          \\
         \save-<1.5em,-1.3em>
         *\txt{Green- \\Schwarz \\ mechanism} \restore
          &  & & SU(3) & \times &  SU(2)_L & \times &  U(1)_Y
          }
         }
\vskip 12pt \centerline{{\bf Figure \the\figno :}\it ~~ Gauge
symmetry breaking of
              $U(4)\times U(2)_L\times U(2)_R$}
\endinsert
\endgroup

It is evident
from Table 13 that there is also something unusually going on
with the right handed leptons. Only one of them
is realized as a bifundamental field, the remaining
two are given by symmetric representations of $U(1)$.
This behavior surely will have consequences for the allowed
couplings, in particular for the Yukawa couplings and the electroweak Higgs
mechanism.

Computing the gauge couplings, we find the following
ratios for the internal volumes of the four 3-cycles
\eqn\ratio{ {\rm Vol(D6_2)}={\rm Vol(D6_4)}=3{\rm Vol(D6_1)}, \quad
             {\rm Vol(D6_3)}=4{\rm Vol(D6_1)} .}
This allows us to determine the ratio of the Standard Model
gauge couplings at the GUT scale to be again
\eqn\ratiob{    {\alpha_{s}\over \alpha_Y}={11\over 3}, \quad\quad
                {\alpha_{w}\over \alpha_Y}={11\over 9} }
leading to a Weinberg angle $\sin^2 \theta_w={9/20}$.
Thus, both models provide  the same prediction for the Weinberg-angle
at the GUT scale.

\bigno
{\it 6.4.3. Electroweak symmetry breaking}
\meno
Finally, we would like to make some comments on electroweak
symmetry breaking in this model.
From the quiver diagram of the $U(4)\times U(2)\times U(2)$ Pati-Salam
model we do not expect that the three Higgs fields in the $(1,\o{2},2)$
representation
get a mass during the brane recombination process. Therefore,
our model does contain appropriate Higgs fields to participate in the electroweak
symmetry breaking.
The three Higgs fields, $H_{bc}$,  in the Pati-Salam model in Table 10
give rise to the Higgs fields
\eqn\higgsfield{  H_{BD}=(1,2)_{(0,1,0,-1)} +c.c.\, ,
                  \quad\quad H_{B'C}=(1,2)_{(0,1,1,0)} +c.c.}
for the $SU(3)_c\times SU(2)_L\times U(1)_Y$ model above.

Of course supersymmetry should already be broken by some mechanism
above the electroweak symmetry breaking scale, but nevertheless
we can safely discuss the expectations  from the purely
topological data of the corresponding brane recombination process.
Since we do not want to break the color $SU(3)$, we still take a
stack of three D6-branes which are wrapped on the cycle $\pi_\alpha=\pi_A$.
Giving a VEV to the fields $H_{BD}$ is expected to correspond
to the brane recombination
\eqn\bewbrane{  \pi_\beta=\pi_B+\pi_D.}
However, for the brane recombination
\eqn\newbrane{
\pi_\gamma=\pi_B+\pi'_C, }
the identification with the corresponding
field theory deformation is slightly more subtle, as the
intersections between these two branes  support both the massless
chiral  multiplet $l_L^{B'C}$ as listed in Table 13 and the Higgs
field $H_{B'C}$. Thus, the intersection preserves only ${\cal
N}=1$ supersymmetry and one might expect that some combination of
$l_L^{B'C}$ and  $H_{B'C}$ are involved in the brane recombination
process. Even without knowing all the details, in the following we
can safely compute the chiral spectrum via intersection numbers.

After the brane recombination we have
a naive gauge group $U(3)\times U(1)\times U(1)$, which however is
broken by the Green-Schwarz couplings to $SU(3)_c\times
U(1)_{em}$ with
\eqn\elektro{   U(1)_{em}={1\over 6}
U(1)_\alpha-{1\over 2} U(1)_\beta +
                  {1\over 2} U(1)_\gamma.}
Interestingly, just $U(1)_{em}$ survives this brane recombination
process. Moreover, all intersection numbers vanish, so that there
are no  chiral massless fields, i.e. all quark and leptons
in Table 13 have gained a mass including the left-handed neutrinos
and the exotic matter.
Looking at the charges in Table 13, one realizes that in the
leptonic sector this Higgs effect cannot be the usual one, where simply
$l_L$ and $e_R$ receive a mass via some Yukawa couplings.
Here also higher dimensional couplings, like the dimension five coupling
\eqn\higherc{   W\sim  {1\over M_s} \o{H}_{BD}\,\o{H}_{BD}\,  S\, e^{D'D}_R ,}
are relevant. These couplings induce a mixing of the Standard Model
matter with the exotic field, $S$. Thus we can state, that
by realizing some of the right handed leptons in the (anti-)symmetric
representation, the exotic field is needed to give
all leptons a mass during electroweak symmetry breaking.
It remains to be seen whether the induced masses can be consistent
with the low-energy data.

\newsec{Conclusions}

In this paper we have studied intersecting brane worlds for the
$T^6/\ZZ_4$ orientifold background with special emphasis on
supersymmetric configurations. We have found as a first
non-trivial result a globally supersymmetric three generation
Pati-Salam type extension of the Standard Model with some exotic
matter. The chiral matter content is only slightly extended by one
chiral multiplet in the (anti-)symmetric representation of
$SU(2)_L$. The presence of this exotic matter can be traced back
to the fact that we were starting with a Pati-Salam gauge group,
where the anomaly constraints forced us to introduce additional
matter. Issues which arose for non-supersymmetric models will also
appear in the supersymmetric setting. Since the Green-Schwarz
mechanism produces global $U(1)$ symmetries, the allowed couplings
in the effective gauge theory are usually much more constrained
than for the Standard Model.

With such model a hand, many
phenomenological issues deserve to be studied, as for instance
mechanisms for supersymmetry breaking, the generation of soft
breaking terms, Yukawa and higher dimensional couplings, the
generation of $\mu$-terms and gauge coupling unification. It also
remains to be seen whether the electroweak Higgs effect indeed
produces the correct masses for all quarks and leptons. Moreover,
one should check whether the renormalization of the gauge
couplings from the string respectively the PS-breaking scale down
to the weak scale can lead to acceptable values for the Weinberg
angle.

The motivation  for this analysis was to start a
systematic search for realistic supersymmetric intersecting brane
world models. We have worked out some of the technical model
building aspects when one is dealing with more complicated
orbifold backgrounds containing in particular twisted sector
3-cycles. These techniques can be directly generalized to, for
instance, the $\ZZ_6$ orientifolds  \rbgkc\ or the $\ZZ_N\times
\ZZ_M$  orientifold  models \rfhs. It could be worthwhile to
undertake a similar study for these orbifold models, too.

The final goal would be to find a realization of the
MSSM in some simple intersecting brane world model.
As should have become clear from our analysis, while
phenomenologically interesting non-supersymmetric models are fairly
easy to get, the same is not true for the supersymmetric
ones. Requiring supersymmetry imposes very strong constraints
on the possible configurations and as we have observed in the
$\ZZ_4$ example, also the supply of possible  intersection numbers
is very limited.
These obstructions appear to be less surprising, when one contemplates that
for smooth backgrounds, by lifting to M-theory,
the construction of an ${\cal N}=1$
chiral intersecting brane world background with $O6$ planes and
$D6$ branes is equivalent
to the construction of a compact singular $G_2$ manifold.
In this respect it would be interesting whether certain M-theory
orbifold constructions like the one discussed in \rfaux\ are dual to
the kind of models discussed in this paper.

At a certain scale close to the TeV scale supersymmetry has to be
broken. For the intersecting brane world scenario one might
envision different mechanisms for such a breaking. First, we might
use the conventional mechanism of gaugino condensation  via some
non-perturbative.
Alternatively, one could build models where the MSSM is localized on a
number of D-branes, but where the RR-tadpole cancellation
conditions requires the introduction of hidden sector branes, on
which supersymmetry might be broken. This breaking could be
mediated gravitationally to the standard model branes. A third
possibility is to get D-term supersymmetry breaking by generating
effective Fayet-Iliopoulos terms via complex structure
deformations. We think that these issues and other
phenomenological questions deserve to be studied in the future.

\vskip 2cm

\centerline{{\bf Acknowledgments}}\pano

We would like to thank V. Braun, A. Klemm, B. K\"ors, D. L\"ust and
S. Stieberger for
helpful discussion.
This work is supported in part by the EC under the RTN project
HPRN-CT-2000-00131. T.O. also thanks the Graduiertenkolleg {\it
The Standard Model of Particle Physics - structure, precision tests
and extensions} maintained by the DFG.
The work of L.G. is supported by the DFG priority program (1096)
under the project  number DFG Lu 419/7-2.

\vfill\eject
\appendix{A}{Orientifold planes}

In this appendix we present the results for the O6-planes
and the action of $\Omega\o\sigma$ on the homology lattice
for the other three orientifold models.
We have listed the results in Table A1.
\vskip 0.8cm
\vbox{
\centerline{\vbox{
\hbox{\vbox{\offinterlineskip
\def\tablespace{height2pt&\omit&&
 \omit&\cr}
\def\tablerule{\tablespace\noalign{\hrule}\tablespace}

\hrule\halign{&\vrule#&\strut\hskip0.2cm\hfill
#\hfill\hskip0.2cm\cr & model  && O6-plane &\cr \tablerule & ${\bf
AAA}$   && $4\, \rho_1 -2\, \o\rho_2 $   &\cr \tablerule & ${\bf
AAB}$   && $2\, \rho_1 + \rho_2 -2\, \o\rho_2 $   &\cr \tablerule
& ${\bf ABA}$   && $2\, \rho_1 + 2\, \rho_2+  2\, \o\rho_1  -2\,
\o\rho_2 $   &\cr \tablerule & ${\bf ABB}$   && $ 2\, \rho_2+2\,
\o\rho_1 -2\, \o\rho_2 $   &\cr }\hrule}}}} \centerline{
\hbox{{\bf Table A1:}{\it ~~ O6-planes}}} } \vskip 0.5cm \noindent
For the action of $\Omega\o\sigma$ on the orbifold basis we find:

\item{{\bf AAA}:}
For the toroidal 3-cycles we get
\eqn\actbasa{\eqalign{ &\rho_1\to \rho_1, \quad  \o\rho_1\to -\o\rho_1 \cr
                      &\rho_2\to -\rho_2, \quad  \o\rho_2\to \o\rho_2 \cr}}
and for the exceptional cycles
\eqn\actbasexa{\eqalign{ &\varepsilon_i\to -\varepsilon_i \quad\quad
                        \o\varepsilon_i\to \o\varepsilon_i ,\cr}}
for all $i\in\{1,\ldots,6\}$.

\item{{\bf AAB}:}
For the toroidal 3-cycles we get
\eqn\actbasb{\eqalign{ &\rho_1\to \rho_1, \quad  \o\rho_1\to \rho_1 -\o\rho_1 \cr
                      &\rho_2\to -\rho_2, \quad  \o\rho_2\to -\rho_2+\o\rho_2 \cr}}
and for the exceptional cycles
\eqn\actbasexb{\eqalign{ &\varepsilon_i\to -\varepsilon_i \quad\quad
                        \o\varepsilon_i\to -\varepsilon_i+\o\varepsilon_i ,\cr}}
for all $i\in\{1,\ldots,6\}$.

\item{{\bf ABA}:}
For the toroidal 3-cycles we get
\eqn\actbasc{\eqalign{ &\rho_1\to \rho_2, \quad  \o\rho_1\to -\o\rho_2 \cr
                      &\rho_2\to \rho_1, \quad  \o\rho_2\to -\o\rho_1 \cr}}
and for the exceptional cycles
\eqn\actbasexc{\eqalign{ &\varepsilon_1\to \varepsilon_1 \phantom{-}\quad\quad
                        \o\varepsilon_1\to -\o\varepsilon_1 \cr
                        &\varepsilon_2\to  \varepsilon_2 \phantom{-} \quad\quad
                        \o\varepsilon_2\to -\o\varepsilon_2 \cr
                        &\varepsilon_3\to -\varepsilon_3 \quad\quad
                        \o\varepsilon_3\to \o\varepsilon_3 \cr
                        &\varepsilon_4\to -\varepsilon_4 \quad\quad
                        \o\varepsilon_4\to \o\varepsilon_4 \cr
                        &\varepsilon_5\to -\varepsilon_6  \quad\quad
                        \o\varepsilon_5\to \o\varepsilon_6 \cr
                        &\varepsilon_6\to -\varepsilon_5 \quad\quad
                        \o\varepsilon_6\to \o\varepsilon_5 .\cr}}

\appendix{B}{Supersymmetry conditions}

\noindent
In this appendix we list the supersymmetry conditions for the remaining
three orientifold models.

\item{{\bf AAA}:} The condition that such a D6-brane preserves the same supersymmetry
as the orientifold plane is simply
\eqn\susya{      \varphi_{a,1}+ \varphi_{a,2}+ \varphi_{a,3}=0
                 \ {\rm mod}\ 2\pi}
with
\eqn\tangia{  \tan\varphi_{a,1}={m_{a,1}\over n_{a,1}}, \quad
              \tan\varphi_{a,2}={m_{a,2}\over n_{a,2}}, \quad
               \tan\varphi_{a,3}={U_2\, m_{a,3}\over n_{a,3} } .}
This implies the following necessary condition in terms of the wrapping
numbers
\eqn\susywara{  U_2=-{ n_{a,3}  \over m_{a,3}}
                  { \left(n_{a,1}\,m_{a,2} +
               m_{a,1}\,n_{a,2} \right)
           \over \left(n_{a,1}\,n_{a,2} - m_{a,1}\,m_{a,2}  \right)}. }

\item{{\bf AAB}:} The condition that such a D6-brane preserves the same supersymmetry
as the orientifold plane is simply
\eqn\susyb{      \varphi_{a,1}+ \varphi_{a,2}+ \varphi_{a,3}=0
                 \ {\rm mod}\ 2\pi}
with
\eqn\tangib{  \tan\varphi_{a,1}={m_{a,1}\over n_{a,1}}, \quad
              \tan\varphi_{a,2}={m_{a,2}\over n_{a,2}}, \quad
               \tan\varphi_{a,3}={U_2\, m_{a,3}\over n_{a,3}+{1\over 2} m_{a,3} } .}
This implies the following necessary condition in terms of the wrapping
numbers
\eqn\susywarb{  U_2=-{ \left(n_{a,3}+{1\over 2} m_{a,3}\right)  \over m_{a,3}}
                  { \left(n_{a,1}\,m_{a,2} +
               m_{a,1}\,n_{a,2} \right)
           \over \left(n_{a,1}\,n_{a,2} - m_{a,1}\,m_{a,2}  \right)}. }

\item{{\bf ABA}:} The condition that such a D6-brane preserves the same supersymmetry
as the orientifold plane is simply
\eqn\susyc{      \varphi_{a,1}+ \varphi_{a,2}+ \varphi_{a,3}={\pi\over 4}
                 \ {\rm mod}\ 2\pi}
with
\eqn\tangic{  \tan\varphi_{a,1}={m_{a,1}\over n_{a,1}}, \quad
              \tan\varphi_{a,2}={m_{a,2}\over n_{a,2}}, \quad
               \tan\varphi_{a,3}={U_2\, m_{a,3}\over n_{a,3} } .}
This implies the following necessary condition in terms of the wrapping
numbers
\eqn\susywarc{  U_2={ n_{a,3} \over m_{a,3}}
                  { \left(n_{a,1}\,n_{a,2} - m_{a,1}\,m_{a,2} - n_{a,1}\,m_{a,2} -
               m_{a,1}\,n_{a,2} \right)
           \over \left(n_{a,1}\, n_{a,2} - m_{a,1}\,m_{a,2} + n_{a,1}\,m_{a,2} +
            m_{a,1}\,n_{a,2} \right)}. }

\appendix{C}{Fractional boundary states}

\noindent
The unnormalized boundary states in light cone gauge for D6-branes at angles in the
untwisted sector are given by
\eqn\bounda{\eqalign{ |D;(n_I,m_I)\rangle_{U}=
                &|D;(n_I,m_I),NSNS,\eta=1\rangle_{U}+
                   |D;(n_I,m_I),NSNS,\eta=-1\rangle_{U}+ \cr
                   &|D;(n_I,m_I),RR,\eta=1\rangle_{U}+
                   |D;(n_I,m_I),RR,\eta=-1\rangle_{U} }}
with the coherent state
\eqn\boundb{\eqalign{
|D;(n_I,m_I),\eta\rangle
   = \int dk_2 dk_3 \sum_{\vec r,\vec s} {\rm exp}
 \biggl(&-\sum_{\mu=2}^3 \sum_{n>0} {1\over n} \alpha^\mu_{-n}
               \tilde \alpha^\mu_{-n} \cr
       &-\sum_{I=1}^3    \sum_{n>0} {1\over 2n} \left(e^{2i\varphi_I}
 \zeta^I_{-n} \tilde{\zeta}^I_{-n}  +
  e^{-2i\varphi_I} \o\zeta^I_{-n} \tilde{\o\zeta}^I_{-n} \right)\cr
       &+ i \eta \bigl[ {\rm fermions} \bigr] \biggr)
             |\vec r,\vec s,\vec k, \eta \rangle\, .\cr}}
Here $\alpha^\mu$ denotes the two real non-compact directions and
$\zeta^I$  the three complex compact directions.
The angles $\varphi_I$ of the D6-brane relative to the horizontal
axis on each of the three internal tori $T^2$ can be expressed
by the wrapping numbers $(n_I,m_I)$ as listed in Appendix B.
The boundary state \boundb\ involves a sum over the internal
Kaluza-Klein and winding ground states parameterized by
$(\vec r,\vec s)$.
The mass of  these KK and winding modes on each $T^2$ in general reads
\eqn\spec{  M^2_{I}={ |r_I+s_I\,{U_I}|^2\over U_{I,2} }\,
                     { |n_I+m_I\, {T_{I}}|^2\over {T_{I,2}} }}
with $r_I,s_I\in\ZZ$ as above and $U_I$ and $T_I$ denote the
complex and K\"ahler structure on the torus \rbgkl. If the brane
carries some discrete Wilson lines, $\vartheta=1/2$,  appropriate
factors of the form $e^{i s R \vartheta}$ have to be introduced
into the winding sum in \boundb.

\noindent In the $\Theta^2$ twisted sector, the boundary state
involves the analogous sum over the fermionic spin structures
\bounda\ with \eqn\boundc{\eqalign{|D;(n_I,m_I),e_{ij}, \eta
                  \bigr\rangle_T=\int dk_2 dk_3  \sum_{r_3,s_3} {\rm exp}
 \biggl(&-\sum_{\mu=2}^3 \sum_{n>0} {1\over n} \alpha^\mu_{-n}
               \tilde \alpha^\mu_{-n}  \cr
    &-\sum_{I=1}^2    \sum_{r\in\ZZ^+_0 +{1\over 2}} {1\over 2r} \left(e^{2i\varphi_I}
 \zeta^I_{-r} \tilde{\zeta}^I_{-r}  +
  e^{-2i\varphi_I} \o\zeta^I_{-r} \tilde{\o\zeta}^I_{-r} \right)\cr
    &-\sum_{n>0 } {1\over 2n} \left (e^{2i\varphi_3}
 \zeta^3_{-n} \tilde{\zeta}^3_{-n}  +
  e^{-2i\varphi_3} \o\zeta^3_{-n} \tilde{\o\zeta}^3_{-n} \right)\cr
       &+ i \eta \bigl[ {\rm fermions} \bigr] \biggr)
             |r_3,s_3,\vec k,e_{ij},\eta \rangle\, .\cr}}
where $e_{ij}$ denote the 16 $\ZZ_2$ fixed points. Here, we have
taken into account that the twisted boundary state can only have
KK and winding modes on the third $T^2$ torus and that the bosonic
modes on the two other $T^2$ tori carry half-integer modes.

\vfill\eject

\listrefs

\bye
\end